\documentclass[aps,prb,
preprintnumbers,amsmath,amssymb
]{revtex4-2}

\usepackage{graphicx}
\usepackage{dcolumn}
\usepackage{bm}

\usepackage{epsfig}
\usepackage{latexsym}
\usepackage{color}
\usepackage{epstopdf,mathtools,braket}
\numberwithin{equation}{section}
\allowdisplaybreaks
\usepackage{caption}
\usepackage{subcaption}
\usepackage{url}
\usepackage{comment}

\usepackage{graphicx}
\def\be{\begin{equation}}
\def\ee{\end{equation}}
\def\bea{\begin{eqnarray}}
\def\eea{\end{eqnarray}}
\def\bequ{\begin{equation}}
\def\eequ{\end{equation}}

\newcommand{\hx}{\hat{x}}
\newcommand{\hy}{\hat{y}}
\newcommand{\ha}{\hat{a}}
\newcommand{\hb}{\hat{b}}

\def\be{\begin{equation}}
\def\ee{\end{equation}}
\def\bea{\begin{eqnarray}}
\def\eea{\end{eqnarray}}

\begin{document}
\preprint{YITP-23-129, RIKEN-iTHEMS-Report-23}

\title{Foliated BF theories and Multipole symmetries}

\author{Hiromi Ebisu$^{1,2}$}
\email{hiromi.ebisu(at)yukawa.kyoto-u.ac.jp}

\author{Masazumi Honda$^{2,3,4}$}
\email{masazumi.honda(at)yukawa.kyoto-u.ac.jp}

\author{Taiichi Nakanishi$^{2,3}$}
\email{taiichi.nakanishi(at)yukawa.kyoto-u.ac.jp}

\affiliation{$^{1}$Department of Physics and Astronomy, Rutgers University, Piscataway, NJ 08854, USA}
\affiliation{$^{2}$Center for Gravitational Physics and Quantum Information, Yukawa Institute for Theoretical Physics, Kyoto University, Sakyo-ku, Kyoto 606-8502, Japan}
\affiliation{$^{3}$Interdisciplinary Theoretical and Mathematical Sciences Program (iTHEMS),\\ RIKEN, Wako 351-0198, Japan}
\affiliation{$^{4}$Theoretical Sciences Visiting Program, Okinawa Institute of Science and Technology Graduate University (OIST), Onna, 904-0495, Japan}


\date{\today}


\begin{abstract}
Due to the recent studies of the fracton topological phases, which host deconfined quasi-particle excitations with mobility restrictions, the concept of symmetries have been updated. 
Focusing on one of such new symmetries, multipole symmetries, including global, dipole, and quadruple symmetries, and gauge fields associated with them, we construct a new sets of $\mathbb{Z}_N$ $2+1d$ foliated BF theories, where BF theories of conventional topological phases are stacked in layers with couplings between them. 
By investigating gauge invariant non-local operators, we show that our foliated BF theories exhibit unusual ground state degeneracy depending on the system size; it depends on the greatest common divisor between $N$ and the system size. 
Our result provides a unified insight on UV lattice models of the fracton topological phases and other unconventional ones in view of foliated field theories.
\end{abstract}

\maketitle




\noindent

\section{Introduction}
Topologically ordered phases
are unconventional phases of matter and
have been one of the central subjects in
condensed matter physics community~\cite{Tsui,Laughlin1983,Kalmeyer1987,wen1989chiral,Wen:1989iv,PhysRevLett.66.1773}. The prominent feature of these phases is that they host
exotic fractionalized excitations, called anyons~\cite{leinaas1977theory,Laughlin1983,anyon_PhysRevLett.49.957}. These excitations are not only theoretically intriguing but also may find potential application in quantum information science as exchanging these excitations can be utilized for quantum computers~\cite{dennis2002topological,KITAEV20032}.
\par
Recently, new types of topologically ordered phases have been introduced, referred to as the fracton topological phases~\cite{Chamon:2004lew,BRAVYI2011839,Haah:2011drr,Vijay:2015mka,Nandkishore:2018sel,Pretko:2020cko}. 
The distinctive feature of these phases
is that 
a mobility constraint is imposed on quasi-particle excitations, 
leading to the sub-extensive ground state degeneracy~(GSD). 
Due to this feature, conventional effective field theory description of the topologically ordered phases~\cite{witten1989quantum,Elitzur:1989nr,Blau:1989bq,wen2004quantum} cannot be applied to the fracton topological phases.
Fractons have attracted a lot of interests 
in the the field of high energy physics. 
Indeed, fractons have been recently studied in the context of the gravity theory~\cite{Pretko:2017fbf,Benedetti:2021lxj,Benedetti:2022zbb,Hinterbichler:2022agn}, the branes~\cite{Geng:2021cmq}, and holography~\cite{Yan:2018nco,Yan:2019quy,Yan:2019vzp}.
Given the novelty of these phases, 
one of the challenges is to establish a consistent framework for continuum field theories.\par
One of the attempts to tackle this problem is to introduce new types of symmetries, \textit{sub-system symmetries} and \textit{multipole symmetries}~\cite{PhysRevB.66.054526,Seiberg:2019vrp,griffin2015scalar,Pretko:2018jbi,PhysRevX.9.031035,Jain:2021ibh,bidussi2022fractons}.
Here, sub-system symmetry means that a theory is invariant under a symmetry action which acts on a sub-manifold.
The multipole symmetry, especially, the $U(1)$ multipole symmetry, is the generalization of the global $U(1)$ symmetry in the sense that a theory is invariant under the global phase rotation which depends on the spatial coordinate in polynomial form. 
For instance,
in the case of a scalar theory which respects the global and dipole $U(1)$ symmetries, the Lagrangian is invariant under $\Phi\to e^{i\alpha+i\beta x}\Phi$, where $\alpha$ and $\beta$ are constants, and $x$ denotes the spatial coordinate~\cite{Pretko:2018jbi}. 
Investigating the fracton topological phases in view of such new type of symmetries have been recently started~\cite{You:2018bmf,Tantivasadakarn:2020lhq,Shirley:2020ass,Seiberg:2020bhn,Seiberg:2020wsg,Seiberg:2020cxy,Gorantla:2020xap,Gorantla:2020jpy,Rudelius:2020kta,Gorantla:2021svj,Gorantla:2021bda,Yamaguchi:2021qrx,Razamat:2021jkx,Geng:2021cmq,Distler:2021qzc,Burnell:2021reh,Yamaguchi:2021xeq,Katsura:2022xkg,2022arXiv220907987E,Gorantla:2022ssr,Gorantla:2022pii,Cao:2022lig,honda2022scalar,Cao:2023doz,Cao:2023rrb}.

Another strategy to construct effective field theories of the fracton topological phases is to introduce the so-called foliated BF theories~\cite{foliated1,foliated2,foliated3,spieler2023exotic,Ohmori:2022rzz}. 
Such theories are introduced so that 
2+1d BF theories are stacked in layers.
Important aspect of these theories is that the gauge transformation is modified due to the couplings between the layers so that a constraint is imposed on the form of the gauge invariant operators, such as the Wilson loops,
contributing to the sub-extensive GSD. To our knowledge, there are handful number of foliated BF theories that are known to describe the fracton topological phases, such as the~$X$-cube model. Complete understanding of the foliated BF theories has yet to be elusive.

In this work, we explore new sets of 2+1d foliated BF theories 
by 
taking one of the new type of symmetries, multipole symmetries into account. Introducing gauge fields associated with the multipole symmetries, one can systematically construct new foliated BF theories. We further show that such theories exhibit the unusual GSD dependence on the system size; the GSD depends on the greatest common divisor between a quantum number~$N$ characterizing the fractional charges and the system size, which is in contrast with the previous foliated BF theories which show the sub-extensive GSD~\cite{foliated1,foliated2}. We also show that our foliated BF theories have the UV lattice model counterparts which were studied in the literature~\cite{pace_Wen_2022,hotan2022,PhysRevB.107.125154}, sometimes referred to as the higher rank topological phases. 
Consideration given in this work provides a unified insight on the fracton topological phases and the higher rank topological phases~\cite{pace_Wen_2022,hotan2022,PhysRevB.107.125154,guilherme2023,oh2023aspects} in view of foliated field theories.\par
The rest of this work is organized as follows. In Sec.~\ref{sec2}, to see clearly how our strategy of building up the foliated BF theories works, we discuss a simple example of construction of a BF theory of a conventional topological phase, starting with a theory with a global $U(1)$ symmetry. In Sec.~\ref{sec3}, 
 we study a foliated BF theory with global and dipole symmetry. We also discuss the GSD of the theory on a torus geometry and the UV lattice model counterpart. In Sec.~\ref{sec4}, we further explore a foliated BF theory with global, dipole, and quadrupole symmetry. 
 Finally, in Sec.~\ref{sec5}, we conclude our work with a few future directions. Technical issues are relegated into the appendices.

\section{Warm-up: construction of the $\mathbb{Z}_N$ toric code}
\label{sec2}

Before going into detailed discussion of the foliated BF theories, 
we demonstrate a way to
construct the BF theory of the $\mathbb{Z}_N$ toric code~\cite{KITAEV20032} starting with a theory with global $U(1)$ symmetry and the corresponding $U(1)$ gauge field for clearer illustrations. 
Throughout this paper, we study theories in $2+1$ dimensions. 
Also, we employ the differential form for notational simplicity.
\par
We consider
a theory with global $U(1)$ $0$-form symmetry and its charge $Q$.
The conserved charge is described by 
\begin{equation}
    Q(V)=\int_V * j, 
\end{equation}
where $j$ and $V$ represents the conserved 1-form current and two dimensional spatial volume, and $*$ does the Hodge dual.
This charge is global in the sense that it commutes with the translation operation, $P_{I}\;(I=x,y)$ i.e., 
\begin{equation}
   \left[ P_I,Q \right] =0
 \label{eq:re}.
\end{equation}
We introduce a $1$-form $U(1)$ gauge field $a$ which couples with the current $j$ with the coupling term being described by 
\begin{eqnarray}
    S_{c}=\int_Va\wedge* j\label{cp}.
\end{eqnarray}
With the gauge transformation ($\chi$: gauge parameter)
   $ a\ \to\  a+d\chi$
and the condition that the coupling term~\eqref{cp} is gauge invariant, we have the conservation law of the current $d* j=0$.
Defining a gauge invariant flux as $f\vcentcolon=da$, we introduce the following Lagrangian:
\begin{equation}
    \mathcal{{L}}_{TC}=\frac{N}{2\pi}b\wedge f=\frac{N}{2\pi}b\wedge da, \label{toric code}
\end{equation}
where $b$ represents a $1$-form $U(1)$ gauge field.
The theory~\eqref{toric code} is 
nothing but the BF description of the~$\mathbb{Z}_N$ toric code~\cite{KITAEV20032,hansson2004superconductors}.
Equation of motion of the theory~\eqref{toric code} implies that the following gauge invariant field strength vanish
\begin{eqnarray}
  && B^a=\partial_xa_y-\partial_ya_x,\quad      E^a_x=\partial_{\tau}a_x-\partial_xa_0,\quad E^a_y=\partial_{\tau}a_y-\partial_ya_0 ,\nonumber\\
   &&  B^b=\partial_xb_y-\partial_yb_x,\quad  E^b_x=\partial_{\tau}b_x-\partial_xb_0,\quad E^b_y=\partial_{\tau}b_y-\partial_yb_0.
    \label{gaugetc}
\end{eqnarray}
\par
It is known that 
the BF theory~\eqref{toric code} has non-trivial ground state degeneracy (GSD) when we put the theory on Riemann surface with non-zero genus ~\cite{KITAEV20032,Elitzur:1989nr,hansson2004superconductors}. 
Indeed, on the torus geometry, 
the number of distinct non-contractible loops of the gauge fields, $a_x$, and $a_y$ amounts to be the GSD. 
To see this, we consider the following Wilson loop
\begin{equation*}
   W_{0x}(y)=\exp\biggl[i\oint dx a_x(x,y)\biggr],
\end{equation*}
 which can be intuitively understood as the trajectory of the $\mathbb{Z}_N$ fractional charge going around the torus in the $x$-direction. Since the magnetic flux vanishes, this loop does not depend on $y$. Likewise, we also think of the following Wilson loop
 \begin{equation*}
     W_{0y}(x)=\exp\biggl[i\oint dy a_y(x,y)\biggr],
\end{equation*}
 which is associated with the trajectory of the $\mathbb{Z}_N$ fractional charge in the $y$-direction. 
Using the flux-less condition, one can verify that it does not depend on $x$. 
From $  W^N_{0x}(y)=  W^N_{0y}(x)=1$, it follows that there are $N^2$ distinct Wilson loops of the gauge field $a$, implying that the GSD is given by $N^2$. 
The way we evaluate the GSD on the torus 
geometry 
by counting distinct number of the Wilson loops will be frequently used in the case of the foliated BF theories discussed in the subsequent sections.\par

\begin{figure}[t]
  \begin{center}
        \includegraphics[width=0.3\textwidth]{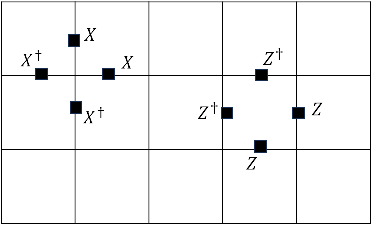}
      \end{center}
      \caption{Two types of the terms defined in~\eqref{tc1}. The black dot represents the $\mathbb{Z}_N$ Pauli operator.
 }
       \label{plane11}
  \end{figure}

One can construct the UV Hamiltonian corresponding to the BF theory~\eqref{toric code}.
Focusing on the field strength of the gauge field $a$, that is, $B^a$, $E^a_x$ and $E^a_y$ in \eqref{gaugetc}, the Hamiltonian of the lattice can be constructed in a way that the ground state does not have the electric and magnetic charges, which are probed by the Gauss law $G=\partial_xE^a_x+\partial_yE^a_y$ and the magnetic flux $B^a$ respectively.
To implement this construction, we think of a 2d discrete lattice and on each link we define a local Hilbert space spanned by
\be
\ket{\eta}_{(\hx+1/2,\hy)} \quad {\rm and}\quad \ket{\eta}_{(\hx,\hy+1/2)} ,
\ee
where the subscripts $(\hx+1/2,\hy)$ and $(\hx,\hy+1/2)$ denote the coordinate of the horizontal and vertical links, respectively while $\eta$ labels $\mathbb{Z}_N$
i.e., $\eta=0,1\cdots,N-1$ $({\rm mod }\ N )$. 
We also introduce the $\mathbb{Z}_N$ Pauli operators $X_{(\hx+1/2,\hy)}$ and $Z_{(\hx+1/2,\hy)}$ acting on the state on the horizontal link $\ket{\eta}_{(\hx+1/2,\hy)}$ as
\begin{eqnarray}
 X_{(\hx+1/2,\hy)}  \ket{\eta}_{(\hx+1/2,\hy)} &=& \ket{\eta+1}_{(\hx+1/2,\hy)} , \nonumber\\
   Z_{(\hx+1/2,\hy)}  \ket{\eta}_{(\hx+1/2,\hy)} &=& \omega^{\eta}\ket{\eta}_{(\hx+1/2,\hy)}, 
\end{eqnarray}
with $\omega\vcentcolon=e^{2\pi i/N}$.
We also define the $\mathbb{Z}_N$ Pauli operators $X_{(\hx,\hy+1/2)}$ and $Z_{(\hx,\hy+1/2)}$ acting on the state on the vertical link in a similar way. 
Using these Pauli operators, we define the following two types of the operators (Fig.~\ref{plane11}):
\begin{eqnarray}
   V_{(\hx,\hy)} &=& X^{\dagger}_{(\hx-1/2,\hy)}X_{(\hx+1/2,\hy)}X^{\dagger}_{(\hx,\hy-1/2)}X_{(\hx,\hy+1/2)} , \nonumber\\     
   P_{(\hx,\hy)} &=& Z^{\dagger}_{(\hx,\hy+1/2)}Z_{(\hx+1,\hy+1/2)}Z^{\dagger}_{(\hx+1/2,\hy)}Z_{(\hx+1/2,\hy+1)} ,\nonumber\\
   &&
\label{tc1}
\end{eqnarray}
which correspond to the Gauss law and the magnetic flux, respectively. 
Then we introduce the Hamiltonian as \cite{KITAEV20032}
\begin{equation}
    H_{TC}=-\sum_{\hx,\hy}\left[ V_{(\hx,\hy)}+ P_{(\hx,\hy)} \right] 
    +({\rm h.c.}).
    \label{tc3}
\end{equation}
The ground state~$\ket{\Omega}$ of this Hamiltonian is the projected state so that it does not have electric and magnetic excitations:
\be
 V_{(\hx,\hy)}\ket{\Omega}=P_{(\hx,\hy)}\ket{\Omega}=\ket{\Omega} \quad
{\rm for}~~^\forall\;\hx,\hy ,
\ee
and the number of the ground states on the torus agrees with the one of the BF theory \eqref{toric code}.
\par
In what follows, we will discuss a way to construct foliated BF theories, starting with a theory with \textit{multipole} symmetries as well as global one. The spirit of obtaining such phases is the essentially the same as what we have discussed here.
namely, starting with conserved charges associated with multipole symmetries, and introducing gauge fields related to them, we define gauge invariant fluxes to introduce the BF theory. 
We also discuss the UV lattice Hamiltonian in the same way as we describe the Hamiltonian~\eqref{tc3} from the BF theory.
The crucial difference between the present argument and what we will see in the following is that due to the multipole symmetries, we obtain richer foliated BF theories.

\section{Dipole symmetries}
\label{sec3}

We turn to the case with global and dipole symmetries. To this end, discussion presented in~\cite{PhysRevB.106.045112,Hirono:2022dci} is of usefulness.
We also discuss the relation between the foliated BF theory and the UV lattice model. 
Suppose that we have a theory with conserved charges associated with global $U(1)$ and dipole symmetries, described by $Q$, $Q_{x}$ and $Q_y$. 
These charges are subject to the following relations
:
\begin{equation}
   \left[ P_I,Q \right] =0 ,\quad 
   \left[ P_I,Q_J \right]=\delta_{I,J}Q \qquad (I=x,y).
    \label{eq:re2}
\end{equation}
where the first relation is the same as \eqref{eq:re}.
We write the charges $Q$, $Q_x$ and $Q_y$ via the integral expression using the conserved currents as
\begin{equation*}
   Q=\int_V*j,\quad Q_I=\int_V*K_I.
\end{equation*}
To reproduce the relation~\eqref{eq:re2}, we demand that 
 \begin{equation}
    *K_I=*k_I-x_I*j
\end{equation}
with $k_I$ being a local (not necessarily conserved) current and $(x_1 ,x_2 )=(x, y )$. 
A straightforward calculation verifies the relation~\eqref{eq:re2}.
Corresponding to~\eqref{cp}, we introduce the $U(1)$ $1$-form gauge fields~$a$ and $A^I$ with the coupling term defined by
\begin{equation}
    S_{dip}=\int_V \left( a\wedge *j+A^I\wedge *k_I \right) .
\end{equation}
We need to have a proper gauge transformation in such a way that the condition of the coupling term being
gauge invariant yields the conservation law of the currents.
The following gauge transformation does the job:
\begin{equation}
    a\to a+d\Lambda+\sigma_Idx_I,\quad A^I\to A^I+d\sigma_I ,
\label{gaugetr1}
\end{equation}
where $\Lambda$, $\sigma_I$ are the gauge parameters.
Indeed, one can verify that the gauge invariance of the coupling term $ S_{dip}$ under the gauge transformation~\eqref{gaugetr1} yields
\begin{equation*}
   d*j=0,\quad  d\left( *k_I-x_I*j \right) =d*K_I=0.
\end{equation*}
\par
Now we are in a good place to introduce the foliated BF theory. Defining gauge invariant fluxes as 
\begin{equation}
    f\vcentcolon=da-A^I\wedge dx_I,\quad F^I\vcentcolon=dA^I,
\end{equation}
we put these fluxes in the BF theory format, namely, we introduce the following BF theory:
\begin{equation}
    \mathcal{L}_{dip}=\frac{N}{2\pi}b\wedge f+\sum_{I=x,y}\frac{N}{2\pi}c^I\wedge F^I,\label{hh}
\end{equation}
where $b$ and $c^I(I=x,y)$ represent the $U(1)$ $1$-form gauge fields. 
To proceed, we also introduce the foliation fields $e^I$ ($I=x,y$)~\cite{foliated1,foliated2}. 
Generally, foliation is defined to be co-dimension one sub-manifold which is orthogonal to the $1$-form foliation field~$e^I$.
Setting the foliation field by $e^x\vcentcolon=dx,e^y\vcentcolon=dy$,  
and rewriting~\eqref{hh}, we arrive at the following foliated BF theory:
\begin{equation}
   \boxed{\mathcal{L}_{dip}=\frac{N}{2\pi}a\wedge db+\sum_{I=x,y}\frac{N}{2\pi}A^I\wedge dc^I+\frac{N}{2\pi}A^I\wedge b\wedge e^I.}\label{foliation dipole}
\end{equation}
Compared with the foliated BF theories that were previously studied~\cite{foliated1,foliated2}, where sub-extensive number of $2+1d$ BF theories are placed in layers, in our theory~\eqref{foliation dipole}, there are only three layers of the $2+1d$ BF theories, corresponding to the first two terms, with coupling between the layers being described by the last term. \par
We study the GSD of our foliated BF theory on a torus. In addition to the gauge symmetry~\eqref{gaugetr1}, the foliated BF theory~\eqref{foliation dipole} admits the following gauge symmetry with respect to $b$ and $c^I$:
\begin{equation}
    b\to b+d\lambda,\quad c^I\to c^I+d\gamma^I+\lambda e^I .
    \label{gaugetr3}
\end{equation}
Similarly to the foliated BF theory of the X-cube model~\cite{foliated1}, the unusual gauge symmetries~\eqref{gaugetr1} and \eqref{gaugetr3} put constraints on the form of the gauge invariant operators, which contributes to the unconventional GSD dependence on the system size. 
To see this point more explicitly, we integrate out some of the fields in~\eqref{foliation dipole} to transform the Lagrangian to a simpler form with fewer number of the gauge fields.\par
Integrating out $b_0$ gives the following condition ($i,j=x,y$):
\begin{equation}
    \partial_{i}a_j\epsilon^{ij}-A^I_i\delta^I_j\epsilon^{ij}=0
    \ 
    \label{eq:eom01}
\end{equation}
We also integrate out the 
$A^I_0$, and obtain the following condition
\begin{equation}
\partial_ic^I_j\epsilon^{ij}+b_i\delta^I_j\epsilon^{ij}=0.
\ 
\label{eq:eom02}
\end{equation}
One can eliminate the gauge fields $b$ and $A^I$ by
substituting the relations~\eqref{eq:eom01}\eqref{eq:eom02} into~\eqref{foliation dipole}.
In doing so, we
introduce gauge fields by 
\begin{equation}
    A_{(ij)}\vcentcolon=\partial_ia_j-A_i^j\quad (i,j=x,y)
    \label{gauge3}
\end{equation}
whose gauge transformation coming from \eqref{gaugetr1} reads 
\begin{eqnarray}
    A_{(ij)}\ \to\  A_{(ij)}+\partial_i\partial_j\Lambda
\end{eqnarray}
with property that $A_{(xy)}=A_{(yx)}$ due to~\eqref{eq:eom01}.
After the substitution, the Lagrangian~\eqref{foliation dipole} becomes

    \begin{eqnarray}
 \mathcal{L}_{dip}
   &=&\frac{N}{2\pi}\biggl[ 
   -c_0^x \left( \partial_xA_{(yx)}-\partial_yA_{(xx)} \right)
   -c_0^y \left( \partial_xA_{(yy)}-\partial_yA_{(xy)} \right)
 +\tilde{c}\left( \partial_{\tau}A_{(xy)}-\partial_x\partial_yA_0 \right) \nonumber\\
  &&
  -c_y^x \left( \partial_{\tau}A_{(xx)}-\partial_x^2A_0 \right)+c_x^y \left( \partial_{\tau}A_{(yy)}-\partial_y^2A_0 \right )\biggr],
  \label{Bfdi}
\end{eqnarray}

where we have defined $A_0\vcentcolon=a_0$ and $\tilde{c}\vcentcolon=c^x_x-c_y^y$ 
with the gauge transformation $\tilde{c}\to \tilde{c}+\partial_x\gamma^x-\partial_y\gamma^y$.

Similarly to the other fracton models, investigating the BF theory~\eqref{Bfdi}, such as evaluation of the form of the Wilson loops and the GSD, is more challenging than the conventional BF theory due to the presence of the higher order spatial derivatives, giving rise to the UV/IR mixing. 
To circumvent this issue, we
follow an approach proposed in~\cite{Gorantla:2021svj} and implement a mapping~\eqref{Bfdi} to what is called ``integer BF theory"~\cite{Gorantla:2021svj,hotan2022,PhysRevB.106.045112} where the gauge fields take integer values defined on a discrete lattice. 
Relegating the details to appendix.~\ref{app1}, we map the theory~\eqref{Bfdi} to the following integer BF theory:
    \begin{eqnarray}
 \mathcal{L}_{dip}
   &=&\frac{2\pi}{N}\biggl[ 
   -\hat{c}_0^x \left( \Delta_x\hat{A}_{(yx)}-\Delta_y\hat{A}_{(xx)} \right)
   -\hat{c}_0^y \left( \Delta_x\hat{A}_{(yy)}-\Delta_y\hat{A}_{(xy)} \right)
 +\hat{\tilde{c}}\left( \Delta_{\tau}\hat{A}_{(xy)}-\Delta_x\Delta_y\hat{A}_0 \right) \nonumber\\
  &&
  -\hat{c}_y^x \left( \Delta_{\tau}\hat{A}_{(xx)}-\Delta_x^2\hat{A}_0 \right)+\hat{c}_x^y \left( \Delta_{\tau}\hat{A}_{(yy)}-\Delta_y^2\hat{A}_0 \right )\biggr],
  \label{Bfdi0}
\end{eqnarray}
where gauge fields with hat~($\hat{\cdot}$) take $\mathbb{Z}_N$ values, which are defined on a discrete lattice. Also, we set the spatial coordinate of the lattice to be $(\hat{x},\hat{y})$ which take integer number in the unit of the lattice spacing, and
define a derivative operator $\Delta_x$ which acts on a function $f$ defined on a site via
$\Delta_x f(\tau,\hx,\hy)=f(\tau,\hx+1,\hy)-f(\tau,\hx,\hy)$ ($\Delta_y,\Delta_\tau$ is similarly defined).
Note that we put hat on top of the gauge fields in~\eqref{Bfdi0} to emphasize that they are integer gauge fields. The gauge fields $A_0$, $c^x_0$, $c^y_0$ reside on the $\tau$-links and $\hat{c}^y_x$ and $\hat{c}^x_y$ do on the $x$ and $y$-links, respectively. Also, $\hat{A}_{(xx)}$ $\hat{A}_{(yy)}$ are defined on sites, whereas $\hat{A}_{xy}$ and $\hat{\tilde{c}}$ are on the plaquettes in the $xy$-plane.
\par
The integer BF theory~\eqref{Bfdi0} admits the following gauge symmetry
\begin{eqnarray}
    \hat{c}_0^x\to \hat{c}_0^x+\Delta_{\tau}\xi^x,\;\hat{c}_y^x\to \hat{c}_0^x+\Delta_{y}\xi^x,
    \hat{c}_0^y\to \hat{c}_0^y+\Delta_{\tau}\xi^y,\;\hat{c}_x^y\to \hat{c}_x^y+\Delta_{x}\xi^y,\hat{\tilde{c}}\to \hat{\tilde{c}}+\Delta_x\xi^x-\Delta_y\xi^y\nonumber\\
\hat{A}_0\to\hat{A}_0+\Delta_{\tau}\eta,\;\hat{A}_{(ij)}\to\hat{A}_{(ij)}+\Delta_{i}\Delta_j\eta,
\end{eqnarray}
where $\xi^x$ $\xi^y$, $\eta$ denote gauge parameters with integer values. Also,
the BF theory~\eqref{Bfdi0} consists of the gauge field $(\hat{c}^x_0,\hat{c}^y_0,\hat{\tilde{c}},\hat{c}^x_y,\hat{c}^y_x)$ and the field strength for $(\hat{A}_0,\hat{A}_{(ij)})$ and vice versa.
The equations of motion of the BF theory implies that the following gauge invariant field strength vanish:
\begin{eqnarray}
 && B_x= \Delta_x \hat{A}_{(yx)}-\Delta_y\hat{A}_{(xx)},\quad  
 B_y= \Delta_x \hat{A}_{(yy)}-\Delta_y\hat{A}_{(xy)} , \nonumber\\
 &&  E_{(ij)}=\Delta_{\tau}\hat{A}_{(ij)}-\Delta_i\Delta_j\hat{A}_0,\nonumber\\
 &&  \tilde{B}=\Delta_x^2 \hat{c}^x_y-\Delta_y^2\hat{c}^y_x-\Delta_x\Delta_y\hat{\tilde{c}},\quad
   E^x_y=\Delta_{\tau}\hat{c}_y^x-\Delta_y\hat{c}^x_0, \nonumber \\
&&
   E^y_x=\Delta_{\tau}\hat{c}_x^y-\Delta_x\hat{c}^y_0,   \tilde{E}=\Delta_{\tau}\hat{\tilde{c}}-\Delta_x\hat{c_0}^x+\Delta_y\hat{c}^y_0. \label{e.o.m}
\end{eqnarray}
It is worth mentioning that the field strength $B_x$, $B_y$, $E_{(xx)}, E_{(yy)},E_{(xy)}$ has the same form as the one which was found in the symmetric tensor gauge theory, unconventional Maxwell theory with the 2-rank spatial-symmetric-tensor space components, admitting dipole charges~\cite{TensorGaugeTheory, Pretko:2016kxt}~\cite{foot3}. 
\par
The equation of motions ensures that there is no non-trivial local gauge invariant operators. 
However, analogously to the other BF theories, 
 the theory has
non-local gauge-invariant operators, which can be constructed from the gauge fields either
$\hat{A}_{(ij)}$ or $(\hat{\tilde{c}},\hat{c}^x_y,\hat{c}^y_x)$. 
Especially, when placing on a torus, the theory admits the non-contractible Wilson loops of the gauge fields which contribute to the non-trivial GSD. 
In the following, we evaluate the GSD of our theory on the torus with system size $L_{x/y}$ in the~$x/y$ direction, by counting the number of distinct non-contractible Wilson loops of the gauge fields $\hat{A}_{(ij)}$~\cite{TechnicalCaveat}.
We set the periodic boundary condition of the lattice via
$(\hx,\hy)\sim(\hx+L_x,\hy)\sim(\hx,\hy+L_y)$.
\par
Let us first focus on the form of the non-contractible Wilson loops of $\hat{A}_{(xx)}$ in the $x$-direction. 
Due to the gauge symmetry~\eqref{gauge3}, there are two types of the loops, described by
\begin{eqnarray}
    W_x(\hy)&=&\exp\biggl[i\frac{2\pi}{N}\sum_{\hx=1}^{L_x}  \hat{A}_{(xx)}(\hx,\hy)\biggr],\nonumber\\  
      W_{dp:x}(\hy)&=&\exp\biggl[i\frac{2\pi}{N}\alpha_x\sum_{\hx=1}^{L_x}  \hat{x}\hat{A}_{(xx)}(\hx,\hy)\biggr],\label{loops}\nonumber\\
    &&
\end{eqnarray}
where $\alpha_x=\frac{N}{\gcd(N,L_x)}$, and $\gcd$ stands for the greatest common divisor.
The first loop in~\eqref{loops} describes the non-contractible loops of the $\mathbb{Z}_N$ fractional charge with~$W_x^N(\hy)=1$, which can be also found in the other topological phases such as the toric code. The second 
loop 
can be interpreted as the ``dipole of the Wilson loop", i.e., the loops is formed by trajectory of the fractional charge with its intensity increasing linearly as going around the torus in the $x$-direction, which is reminiscent of the integration of a dipole moment, $x\rho$. 
These two loops, $W_{x}(\hy)$ and $W_{dp:x}(\hy)$, are characterized by quantum number $\mathbb{Z}_N$ and~$\mathbb{Z}_{\gcd(N,L_x)}$, respectively.
 %
%
We also derive the same Wilson loops from a different perspective, 
see Appendix~\ref{ap:lattice}. \par
Likewise, there are two types of the non-contractible loops of $\hat{A}_{(yy)}$ in the $y$-direction, which has the form
\begin{eqnarray}
      W_y(\hx)=\exp\biggl[i\frac{2\pi}{N}\sum_{\hy=1}^{L_y}  \hat{A}_{(yy)}(\hx,\hy)\biggr],\;
      W_{dp:y}(\hy)=\exp\biggl[i\frac{2\pi}{N}\alpha_y\sum_{\hy=1}^{L_y}  \hy\hat{A}_{(yy)}(\hx,\hy)\biggr]\label{loopsy}
\end{eqnarray}
with  $\alpha_y=\frac{N}{\gcd(N,L_y)}$, implying that these are labeled by $\mathbb{Z}_N\times\mathbb{Z}_{\gcd(N,L_y)}$.
\par
We turn to introducing non-contractible loops of the gauge field~$\hat{A}_{(xy)}$ in either $x$- or~$y$-direction. To this end, we consider the following gauge invariant loops
\begin{eqnarray}
    W_{xy:x}(\hy)&=&\exp\biggl[i\frac{2\pi}{N}\sum_{\hx=1}^{L_x}  \hat{A}_{(xy)}(\hx,\hy)\biggr],\nonumber\\
    W_{xy:y}(\hx)&=&\exp\biggl[i\frac{2\pi}{N}\sum_{\hy=1}^{L_y}  \hat{A}_{(xy)}(\hx,\hy)\biggr].\label{loopsxy}\nonumber\\
    &&
\end{eqnarray}
Naively, there are $N^2$ distinct loops, however, a sets of constraints reduce this number, as we discuss below.\par
After having defined gauge invariant non-contractible loops~\eqref{loops}\eqref{loopsy}\eqref{loopsxy}, we need to check the coordinate dependence of the loops to count the distinct number of the loops. As we will see, there is a several constraints imposed on the loops of the gauge field~$\hat{A}_{(xy)}$.
We first check the $\hy$ dependence of the loops, $W_{x}(\hy)$, $W_{dp:x}(\hy)$~\eqref{loops}.
From the equation of motion of the BF theory~$B_y=0$, where~$B_y$ is given in~\eqref{e.o.m}, and summing the field along the $x$-direction, we have
\begin{equation}
    \sum_{\hx=1}^{L_x}  \left( \Delta_x\hat{A}_{(yx)}-\Delta_y\hat{A}_{(xx)} \right) =0
     \ \leftrightarrow\  \Delta_y \sum_{\hx=1}^{L_x} \hat{A}_{(xx)}=0, 
\end{equation}
implying that the loop $W_{x}(\hy)$~\eqref{loops} does not depend on $\hy$. Furthermore, multiplying $\hx$ on the equation of motion $B_x=0$ and summing along the~$x$-direction, one finds
\begin{equation}
    \Delta_y\sum_{\hx=1}^{L_x}\hx \hat{A}_{(xx)}=-\sum_{\hx=1}^{L_x} \hat{A}_{(xy)},
\end{equation}
from which we have
\begin{equation}
    W_{dp:x}(\hy)=W_{dp:x}(\hy-1)\biggl(W_{xy:x}^{\alpha_x}(\hy)\biggr)^{\dagger}.\label{22}
    \end{equation}
    The iterative use of~\eqref{22} gives
    \begin{equation}
        W_{dp:x}(\hy)=W_{dp:x}(\hy_0)\biggl(W_{xy:x}^{\alpha_x(\hy-\hy_0)}(\hy_0)\biggr)^{\dagger},\label{23}
    \end{equation}
    implying that in order to deform the loop $ W_{dp:x}(\hy)$ from $\hy$ to $\hy_0$, we need to multiply additional loops. 
    Due to the periodic boundary condition, $W_{dp:x}(\hy)=W_{dp:x}(\hy+L_y)$, we must have
\begin{equation}
        W_{xy:x}^{\alpha_xL_y}(\hy)=1.\label{hx}
    \end{equation}
    Therefore, there are $N\times\gcd(N,L_x)$ distinct loops~\eqref{loops} with the condition of~\eqref{hx}.\par
    We turn to checking the $\hx$ dependence of the loops, $W_y(\hx)$, $W_{dp:y}(\hx)$~\eqref{loopsy}.
    The similar line of thoughts leads to that $W_{y}(\hx)$ does not depend on $\hx$, and that 
    \begin{equation}
        W_{dp:y}(\hy)=W_{dp:y}(\hx_0)\biggl(W_{xy:y}^{\alpha_y(\hx-\hx_0)}(\hx_0)\biggr)^{\dagger},\label{234}
    \end{equation}
  from which one finds
    \begin{equation}
          W_{xy:y}^{\alpha_yL_x}(\hx)=1.\label{hy}
    \end{equation}
Hence, there are $N\times\gcd(N,L_y)$ distinct loops~\eqref{loopsy} with the condition of~\eqref{hy}.\par
We move onto the investigating the coordinate dependence of the loops $W_{xy:x}(\hy)$ and $W_{xy:y}(\hx)$ given in \eqref{loopsxy}.
From the equation of motion $B_x=0$, and summing over the field along the $x$-direction, we have
\begin{equation}
    \sum_{\hx=1}^{L_x} (\Delta_x\hat{A}_{(yy)}-\Delta_y\hat{A}_{(xy)})=0\ \leftrightarrow\  \Delta_y\sum_{\hx=1}^{L_x} \hat{A}_{(xy)}=0, 
\end{equation}
from which it can be shown that the loop $ W_{xy:x}(\hy)$ can be deformed so that it can go up or down in the~$y$-direction, namely, $ W_{xy:x}(\hy)= W_{xy:x}(\hy-1)=\cdots=W_{xy:x}(\hy_0)$ where~$\hy_0$ denotes an arbitrary $y$-coordinate (i.e., the loop $ W_{xy:x}(\hy)$ does not depend on $\hy$). Likewise, one can show that the loop $ W_{xy:y}(\hx)$ can go to left or right, i.e, $ W_{xy:y}(\hx)= W_{xy:y}(\hx-1)=\cdots=W_{xy:y}(\hx_0)$ with $\hx_0$ being an arbitrary~$x$-coordinate. Using this property, it follows that
\begin{equation}
   W_{xy:x}^{L_y}(\hy)= W_{xy:y}^{L_x}(\hx).\label{condition1}
\end{equation}
With the conditions~\eqref{hx}, \eqref{hy} and~\eqref{condition1}, one can count the distinct number of the Wilson loops involving $\hat{A}_{(xy)}$~\eqref{loopsxy}. Let us consider a set of loops~\eqref{loopsxy} with the following form
\begin{equation}
     W_{xy:x}^{k_x}(\hy)W_{xy:y}^{k_y}(\hx)\label{comosite}
\end{equation}
with $(k_x,k_y)$ being integers. We need to count the distinct combination of $(k_x,k_y)$. 
Relegating the details to the Appendix~\ref{ap:1}, we find that there are $N\times \gcd(N,L_x,L_y)$ distinct loops of~$\hat{A}_{(xy)}$.\par
To recap the argument, we have counted the distinct number of the non-contractible loops of the gauge fields $\hat{A}_{(ij)}$. 
For the loops involving $\hat{A}_{(xx)}$ and $\hat{A}_{(yy)}$, there are $(N\times\gcd(N,L_x))\times(N\times\gcd(N,L_y))$ distinct loops whereas there are $N\times \gcd(N,L_x,L_y)$ distinct loops of $\hat{A}_{(xy)}$. 
The total number of the distinct loops amounts to be the GSD. Hence, we arrive at
\begin{eqnarray}    \boxed{GSD=N^3\times\gcd(N,L_x)\times\gcd(N,L_y)\times\gcd(N,L_x,L_y).}
\label{dpgsd}
\end{eqnarray}
Compared with previous foliated BF theories~\cite{foliated1,foliated2,foliated3}, where coupling between the layers of the toric codes yields the sub-extensive GSD, in our foliated BF theory, the coupling between the three layers of the toric codes gives rise to unusual GSD dependence on the system size. 

\subsection{UV stabilizer model}
\begin{figure}
    \begin{center}
       \begin{subfigure}[h]{0.20\textwidth}
  \includegraphics[width=\textwidth]{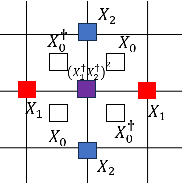}
         \caption{}\label{even00}
             \end{subfigure}
             \hspace{5mm}
               \begin{subfigure}[h]{0.3\textwidth}
    \includegraphics[width=\textwidth]{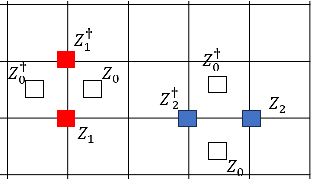}
         \caption{}\label{odd00}
             \end{subfigure} 
 \end{center}
 \caption{Two types of the terms that constitute Hamiltonian~\eqref{op}. The red (blue) square represents the Pauli operator acting on the local Hilbert space~$\ket{a}_{(\hx,\hy)}$ ($\ket{b}_{(\hx,\hy)}$), whereas small square with black line denotes the Pauli operator that act on $\ket{c}_{(\hx+1/2,\hy+1/2)}$. A purple square indicates the composite of the Pauli operators which act on both of $\ket{a}_{(\hx,\hy)}$ and $\ket{b}_{(\hx,\hy)}$.  }
 \end{figure}
 Analogously to the fact that the BF theory description of the $\mathbb{Z}_N$ toric code can be mapped to the stabilizer model in the UV lattice, one can implement the similar mapping from the BF theory to the lattice model. 
 Focusing on gauge invariant field strength, $B_x$, $B_y$, $E_{(ij)}$~\eqref{e.o.m}, 
 we introduce the UV lattice Hamiltonian in such a way that 
 the ground state does not admit the~$\mathbb{Z}_N$ electric and magnetic charge excitations.
\par 
To this end, we envisage a 2d square lattice, and introduce two types of local Hilbert space of the $\mathbb{Z}_N$ spin, denoted by $\ket{a}_{(\hx,\hy)}\ket{b}_{(\hx,\hy)}$ with $a,b=0,1,\cdots,N-1\;\mod N$ as well as~the $\mathbb{Z}_N$ Pauli operators $X_{i,(\hx,\hy)},Z_{i,(\hx,\hy)} (i=1,2)$ that act on the Hilbert space with
\begin{eqnarray}
Z_{1,(\hat{x},\hat{y})}\ket{a}_{(\hat{x},\hat{y})}\ket{b}_{(\hat{x},\hat{y})}=\omega^a\ket{a}_{(\hat{x},\hat{y})}\ket{b}_{(\hat{x},\hat{y})},\quad
Z_{2,(\hat{x},\hat{y})}\ket{a}_{(\hat{x},\hat{y})}\ket{b}_{(\hat{x},\hat{y})}=\omega^b\ket{a}_{(\hat{x},\hat{y})}\ket{b}_{(\hat{x},\hat{y})}\nonumber\\
X_{1,(\hat{x},\hat{y})}\ket{a}_{(\hat{x},\hat{y})}\ket{b}_{(\hat{x},\hat{y})}=\ket{a+1}_{(\hat{x},\hat{y})}\ket{b}_{(\hat{x},\hat{y})},\quad
X_{2,(\hat{x},\hat{y})}\ket{a}_{(\hat{x},\hat{y})}\ket{b}_{(\hat{x},\hat{y})}=\ket{a}_{(\hat{x},\hat{y})}\ket{b+1}_{(\hat{x},\hat{y})}.
\end{eqnarray}
At each node of the dual lattice, labeled by the coordinate~($\hx+1/2,\hy+1/2$), we also introduce a local Hilbert space represented by~$\ket{c}_{(\hx+1/2,\hy+1/2)}$ with $c=0,\cdots,N-1\:\mod N
$ and the~$\mathbb{Z}_N$ Pauli operators, $X_{0,(\hx+1/2,\hy+1/2)}$, $Z_{0,(\hx+1/2,\hy+1/2)}$ which act on the state as
\begin{eqnarray}
    Z_{0,(\hat{x}+1/2,\hat{y}+1/2)}\ket{c}_{(\hat{x}+1/2,\hat{y}+1/2)}
    &=& \omega^c\ket{c}_{(\hat{x}+1/2,\hat{y}+1/2)},\nonumber\\
    X_{0,(\hat{x}+1/2,\hat{y}+1/2)}\ket{c}_{(\hat{x}+1/2,\hat{y}+1/2)}
    &=& \ket{c+1}_{(\hat{x}+1/2,\hat{y}+1/2)}.\nonumber\\
    &&
\end{eqnarray}
With these preparations, we define the following mutually commuting operators:
\begin{eqnarray}
    V_{(\hat{x},\hat{y})} 
    &\vcentcolon=& X_{1,(\hat{x}+1,\hat{y})}X_{1,(\hat{x}-1,\hat{y})}(X_{1,(\hat{x},\hat{y})}^{\dagger})^2X_{2,(\hat{x},\hat{y}+1)}X_{2,(\hat{x},\hat{y}-1)}(X_{2,(\hat{x},\hat{y})}^{\dagger})^2\nonumber\\
&&  X_{0,(\hx+1/2,\hy+1/2)}X^{\dagger}_{0,(\hx-1/2,\hy+1/2)}X_{0,(\hx-1/2,\hy-1/2)}X^{\dagger}_{0,(\hx-1/2,\hy-1/2)},\nonumber\\
P_{(\hat{x},\hat{y}+1/2)}
&\vcentcolon=& Z_{1,(\hat{x},\hat{y}+1)}^{\dagger}Z_{1,(\hat{x},\hat{y})}Z_{0,(\hx+1/2,\hy+1/2)}Z^{\dagger}_{0,(\hx-1/2,\hy+1/2)},\nonumber\\
Q_{(\hat{x}+1/2,\hat{y})}
&\vcentcolon=& Z_{2,(\hat{x}+1,\hat{y})}Z^{\dagger}_{2,(\hat{x},\hat{y})}Z_{0,(\hx+1/2,\hy+1/2)}Z^{\dagger}_{0,(\hx+1/2,\hy-1/2)},
\end{eqnarray}
which are portrayed in Fig.~\ref{even00} and \ref{odd00}. 
The operator $V_{(\hx,\hy)}$ corresponds to the Gauss law term involving the eclectic fields, $E_{(ij)}$,  $G=\Delta_x^2E_{(xx)}+\Delta_{y}^2E_{(yy)}+\Delta_x\Delta_y E_{(xy)}$, whereas the operator $P_{(\hat{x},\hat{y}+1/2)}$ and $Q_{(\hat{x}+1/2,\hat{y})}$ is associated with the magnetic flux $B_x$ and $B_y$, respectively~\cite{TensorGaugeTheory}. The Hamiltonian is defined in such a way that the ground state does not contain the electric and magnetic excitations, namely, 
\begin{equation}
    H_{dip}\vcentcolon=
    -\sum_{\hx,\hy} \left[ V_{(\hx,\hy)}+P_{(\hat{x},\hat{y}+1/2)}+Q_{(\hat{x}+1/2,\hat{y})} \right]
    +({\rm h.c.})~ .
    \label{op}
\end{equation}
It is interesting to note that in~\cite{pace_Wen_2022,oh2023aspects}, the same Hamiltonian was obtained by Higgsing the tensor gauge theory~\cite{TensorGaugeTheory}, and that the same GSD as~\eqref{dpgsd} was derived by counting superselection sectors (i.e., the number of anyons).

\section{Quadrupole}
\label{sec4}
After having seen the example of the BF theories with dipoles symmetries, in this section, we extend the previous argument to the case with quadrupole symmetry. 

\subsection{Construction of the foliated BF theory}
We envisage a theory with quadrupole $U(1)$ symmetry in addition to
 the global charge and dipole ones whose charges are denoted as $Q$, $Q_{x}$, $Q_y$, $Q_{xy}$. These charges are subject to
the following commutation relation with the transnational operators
:
\begin{equation}
[P_I,Q]=0,\quad [P_I,Q_J]=\delta_{I,J}Q,\quad    [P_I,Q_{xy}]=Q_{\bar{I}}, \label{eq:algebra}
\end{equation}
where 
\begin{equation*}
    \bar{I}=\begin{cases}
    x & {\rm for}\ I=y \\
    y & {\rm for}\  I=x \end{cases} .
\end{equation*}
We also write the charges via integral expression of $1$-form currents as
\begin{equation}
   Q=\int_V*j,\quad Q_I=\int_V*K_I,\quad   Q_{xy}=\int_V* \ell
   \label{eq:qpole}.
\end{equation}
If we set
\begin{equation}
    *K_I=*k_I-x_I*j, \quad  
    * \ell\vcentcolon=* J+xy* j-x_I* k_{I},\label{qpole2}
\end{equation}
where $k_I$ and $J$ represent non-conserved local current,
then a simple calculation shows that the form~\eqref{eq:qpole} jointly with~\eqref{qpole2} yields the commutation relations~\eqref{eq:algebra}.
We introduce $1$-form $U(1)$ gauge fields as $a$, $A^I$, and $a^\prime$ that are coupled with the currents with coupling term
\begin{eqnarray}
    S_{qp}=\int _Va\wedge * j+A^I\wedge * k_I+a^\prime\wedge * J.
\end{eqnarray}
With the following gauge transformation, ($\Lambda$, $\sigma_I$, $\Lambda^\prime$ :gauge parameters)
\be
     a\to a+d\Lambda+\sigma_Idx_I,\quad
     A^I \to A^I +d\sigma_I+\Lambda^{\prime} dx_{\bar{I}},\quad 
     a^\prime  \to a^\prime+d\Lambda^\prime,
     \label{gaugetr}
\ee
jointly with the condition $S_{qp}$ is invariant under the gauge transformation, 
one can verify the three kinds of currents are conserved, i.e., $d* j=d* K_I=d* \ell =0$.\par
We further introduce gauge invariant fluxes by
\begin{equation}
    f\vcentcolon=da-A^I\wedge dx^I,\quad 
    F^I\vcentcolon=dA^I-a^\prime\wedge dx^{\bar{I}}, \quad
    f^\prime\vcentcolon=da^\prime.
\end{equation}
Using these fluxes, we defined a BF theory as
\begin{equation}
    \mathcal{L}_{qp}=\frac{N}{2\pi}\biggl[b\wedge f+\sum_{I=x,y}c^I\wedge F^I+D\wedge f^\prime\biggl],
    \label{qp}
\end{equation}
where $c^I$ and $D$ denote $U(1)$ $1$-form gauge fields.
Writing the foliated field as $e^I=dx^I$, 
 \eqref{qp} is transformed into the following foliated BF theory:
 \begin{widetext}
     \begin{equation}
    \boxed{\mathcal{L}_{qp}=\frac{N}{2\pi}a\wedge db+\frac{N}{2\pi}a^\prime\wedge dD+\sum_{I=x,y}\frac{N}{2\pi}A^I\wedge dc^I+\frac{N}{2\pi}A^I\wedge b\wedge e^I+\frac{N}{2\pi}a^\prime\wedge c^I\wedge e^{\bar{I}}}.\label{foliation qupole}
\end{equation}
 \end{widetext}

Compared with the previous foliated BF theory~\eqref{foliation dipole}, we now have four layers of the toric codes, corresponding to the first three terms in~\eqref{foliation qupole}, with the coupling terms between the layers being given by the last two terms. In the next subsection, we show that the foliated BF theory exhibits unusual GSD dependence on the system size on torus geometry by studying the non-contractible Wilson loops.

\subsection{Ground state degeneracy}
In addition to~\eqref{gaugetr}, the theory respects the following gauge symmetry:
\begin{equation}
 D\to D+d\lambda+\xi^Ie^{\bar{I}},\quad c^I\to c^I-d\xi^I.\label{gaugetr2}
\end{equation}
Analogously to the previous foliated BF theory~\eqref{foliation dipole}, the theory admits unusual gauge symmetries due to the presence of the couplings between the layers. This puts constraints on the form of the gauge invariant operators, contributing to the lattice dependence of the GSD.
To see this point, we simplify the Lagrangian~\eqref{foliation qupole} by integrating out some of the fields.\par
By integrating out $b_0$, and $a^\prime_0$ we have
\begin{equation}
    \partial_{i}a_j\epsilon^{ij}-A^I_i\delta^I_j\epsilon^{ij}=0\ \leftrightarrow\  \partial_xa_y-A_{x}^y=\partial_ya_x-A_{y}^x ,
    \label{eq:eom1}
\end{equation}
and
\begin{equation}
    \partial_iD_j\epsilon^{ij}+\sum_{I\neq J}c_i^I\delta^J_j\epsilon^{ij}=0 \ \leftrightarrow\  
    c^x_x-c^y_y=-\partial_iD_j\epsilon^{ij} .
    \label{eq:eom2}
\end{equation}
Furthermore, integrating out the gauge field $A^I_0$ and $c^I_0$ gives rise to the following two conditions:
\begin{equation}
     \partial_ic^I_j\epsilon^{ij}+b_i\delta^I_j\epsilon^{ij}=0,\quad \partial_iA^I_j\epsilon^{ij}+a^\prime_i\delta^{\bar{I}}_j\epsilon^{ij}=0,
\end{equation}
which can be rewritten as
\begin{equation}
    \begin{pmatrix} b_x \\ b_y \end{pmatrix}=\begin{pmatrix}   -\partial_ic^y_j\epsilon^{ij} \\  \partial_ic^x_j\epsilon^{ij} \end{pmatrix},\quad   
    \begin{pmatrix} a^\prime_x \\ a^\prime_y \end{pmatrix}=\begin{pmatrix}  \partial_iA^x_j\epsilon^{ij} \\   -\partial_iA^y_j\epsilon^{ij} \end{pmatrix} .
    \label{eq:eom4}
\end{equation}
To proceed, we introduce gauge fields as
\begin{equation*}
    A_{(xx)}\vcentcolon=\partial_xa_x-A_x^x,\quad A_{(yy)}\vcentcolon=\partial_ya_y-A_y^y,
\end{equation*}
and eliminate the gauge fields $a^\prime$ and $b$ by use of the relations~\eqref{eq:eom1}-\eqref{eq:eom4}, the Lagrangian is rewritten as
\begin{equation}
    \boxed{\mathcal{L}_{qp}=\frac{N}{2\pi}\biggl[A_0\left(\partial_x^2B_{(yy)}-\partial_y^2B_{(xx)}\right)+A_{(xx)}\left(\partial_{\tau}B_{(yy)}-\partial_y^2B_0\right)-A_{(yy)}\left(\partial_{\tau}B_{(xx)}-\partial_x^2B_0\right)\biggr],}
    \label{eq:bf}
\end{equation}
where
\be
A_0\vcentcolon=a_0,\quad B_0\vcentcolon=D_0,\quad B_{(xx)}\vcentcolon=\partial_xD_x+c^y_x,\quad
B_{(yy)}\vcentcolon=\partial_yD_y+c^x_y .
\ee
Analogously to the previous section, in order to study the BF theory, we implement a mapping from~\eqref{eq:bf} to the following integer BF theory defined on a discrete lattice (see appendix.~\ref{app1})
\begin{equation}
   \mathcal{L}_{qp}=\frac{2\pi}{N}\biggl[\hat{A}_0\left(\Delta_x^2\hat{B}_{(yy)}-\Delta_y^2\hat{B}_{(xx)}\right)+\hat{A}_{(xx)}\left(\Delta_{\tau}\hat{B}_{(yy)}-\Delta_y^2\hat{B}_0\right)-\hat{A}_{(yy)}\left(\Delta_{\tau}\hat{B}_{(xx)}-\Delta_x^2\hat{B}_0\right)\biggr]
    \label{eq:bf0}
\end{equation}
where
$\hat{B}_{0},\hat{B}_{(xx)},\hat{B}_{(yy)}$, $\hat{A}_{0}$, $\hat{A}_{(xx)}$, $\hat{A}_{(yy)}$
denote the gauge fields which take integer values. The gauge fields~$\hat{A}_0$ and $\hat{B}_0$ are defined on the $\tau$-links, whereas $\hat{A}_{(ii)}$ and $B_{(ii)}\;(i=x,y)$ are on sites. The gauge fields 
respect the following gauge symmetry ($\xi,\xi^\prime$:integer gauge parameters)
\begin{eqnarray}
   && \hat{B}_0\to \hat{B}_0+\Delta_{\tau}\xi,\quad  
    \hat{B}_{(xx)}\to \hat{B}_{(xx)}+\Delta_x^2\xi,\quad 
    \hat{B}_{(yy)}\to \hat{B}_{(yy)}+\Delta_y^2\xi , \nonumber\\
   &&  \hat{A}_0\to \hat{A}_0+\Delta_{\tau}\xi^\prime,\quad 
     \hat{A}_{(xx)}\to \hat{A}_{(xx)}+\Delta_x^2\xi^\prime,\quad 
     \hat{A}_{(yy)}\to \hat{A}_{(yy)}+\Delta_y^2\xi^\prime.
     \label{tc2}
\end{eqnarray}
The theory~\eqref{eq:bf0} reminds us the BF theory description of the $\mathbb{Z}_N$ toric code~\eqref{toric code} with difference being that the spatial derivatives are replaced with the second order. 
The equation of motions in the Lagrangian imply that the following gauge invariant field strength vanish:
\begin{eqnarray}
    F_{A}=\Delta_x^2\hat{A}_{(yy)}-\Delta_y^2\hat{A}_{(yy)},\quad E_{xA}=\Delta_{\tau}\hat{A}_{(xx)}-\Delta_x^2\hat{A}_0,\quad E_{yA}=\Delta_{\tau}\hat{A}_{(yy)}-\Delta_y^2\hat{A}_0,\nonumber\\
    F_{B}=\Delta_x^2\hat{B}_{(yy)}-\Delta_y^2\hat{B}_{(yy)},\quad E_{xB}=\Delta_{\tau}\hat{B}_{(xx)}-\Delta_x^2\hat{B}_0,\quad E_{yB}=\Delta_{\tau}\hat{B}_{(yy)}-\Delta_y^2\hat{B}_0.
\end{eqnarray}
As in the case with the other BF theories, 
the equations of motions ensure that there is no non-trivial local gauge invariant operators, yet, the theory admits non-local
gauge-invariant operators, contributing to the GSD. 
With the simplified Lagrangian~\eqref{eq:bf0}, we evaluate the GSD on the torus geometry by counting the distinct number of the non-contractible Wilson loops of the gauge fields $\hat{A}_{(xx)}$ and $\hat{A}_{(yy)}$. 
Below, similarly to the previous section, we think of the theory~\eqref{eq:bf0} on a discrete 2d lattice with periodic boundary condition and evaluate the GSD~\cite{foot2}.
\par
As for the non-contractible loops of the gauge field, $\hat{A}_{(xx)}$, referring to the gauge transformation~\eqref{tc2}, 
we have two types of such loops in the form of
\begin{equation}
       W_x(\hy)=\exp\biggl[i\frac{2\pi}{N}\sum_{\hx=1}^{L_x}  \hat{A}_{(xx)}(\hx,\hy)\biggr],\quad  
       W_{dp:x}(\hy)=\exp\biggl[i\frac{2\pi}{N}\alpha_x\sum_{\hx=1}^{L_x}  \hx\hat{A}_{(xx)}(\hx,\hy)\biggr]\label{418}
\end{equation}
with $\alpha_x=\frac{N}{\gcd(N,L_x)}$. 
Justification of the existence of these loops is given in Appendix.~\ref{ap:lattice}, where we discuss thoroughly the Wilson loops in the UV lattice model.
In order to count the number of distinct such loops, 
we need to check whether these loops depend on $\hy$. 
Let us first focus on the loop~$  W_x(\hy)$.
From the equation of motion, $F_A=\Delta_x^2\hat{A}_{(yy)}-\Delta_y^2\hat{A}_{(yy)}=0$, and sum over the field along the~$x$-direction, we have
\begin{equation}
    \Delta_y^2\sum_{\hx=1}^{L_x} \hat{A}_{(xx)}(\hx,\hy)=0\ \leftrightarrow\   \Delta_y\biggl(\sum_{\hx=1}^{L_x} \Delta_y\hat{A}_{(xx)}(\hx,\hy)\biggr)=0,
    \label{419}
\end{equation}
from which it follows that $\sum_{\hx=1}^{L_x} \Delta_y\hat{A}_{(xx)}(\hx,\hy)$ is independent of $\hy$. For an arbitrary $\hy_0$, we set
\begin{eqnarray}
     \sum_{\hx=1}^{L_x} \Delta_y\hat{A}_{(xx)}(\hx,\hy)= \sum_{\hx=1}^{L_x} \Delta_y\hat{A}_{(xx)}(\hx,\hy_0),
\end{eqnarray}
which can be rewritten as
\begin{eqnarray}
     \sum_{\hx=1}^{L_x} \hat{A}_{(xx)}(\hx,\hy+1)-\sum_{\hx=1}^{L_x} \hat{A}_{(xx)}(\hx,\hy)=\sum_{\hx=1}^{L_x} \Delta_y\hat{A}_{(xx)}(\hx,\hy_0).\label{421}
\end{eqnarray}
Iterative use of~\eqref{421} gives
\begin{equation}
   \sum_{\hx=1}^{L_x} \hat{A}_{(xx)}(\hx,\hy)= \sum_{\hx=1}^{L_x} \hat{A}_{(xx)}(\hx,\hy_0)+(\hy-\hy_0)\sum_{\hx=1}^{L_x} \Delta_y\hat{A}_{(xx)}(\hx,\hy_0), 
\end{equation}
from which we have
\begin{equation}
    W_x(\hy)=W_x(\hy_0)(\tilde{W}_x(\hy_0))^{(\hy_0-\hy)},\;\;\tilde{W}_x(\hy_0)\vcentcolon=\exp\biggl[i\frac{2\pi}{N}\sum_{\hx=1}^{L_x} \hat{A}_{(xx)}(\hx,\hy_0+1)-\hat{A}_{(xx)}(\hx,\hy_0)\biggr],\label{diloops}
\end{equation}
indicating that in order to deform the loop $ W_x(\hy)$ from $\hy$ to $\hy_0$, we need to multiply other loops.
The relation~\eqref{diloops} is corroborated by a close investigation of the stabilizer model presented in Appendix.~\ref{ap:lattice}.
\par
The distinct loops of $ W_x(\hy)$ is labeled by two quantum numbers, corresponding to two loops, $W_x(\hy_0)$ and $\tilde{W}_x(\hy_0)$. 
The loop~$W_x(\hy_0)$ is labeled by $\mathbb{Z}_N$ as we have $W_x^N(\hy_0)=1$. Due to the periodic boundary condition, $ W_x(\hy+L_y)= W_x(\hy)$ and~\eqref{diloops}, the loop $\tilde{W}_x(\hy_0)$ is subject to $\tilde{W}_x^{L_y}(\hy_0)=\tilde{W}_x^N(\hy_0)=1$, from which it follows that the loop $\tilde{W}_x(\hy_0)$ is labeled by $\mathbb{Z}_{\gcd(N,L_y)}$. In total, the distinct number of the loop $  W_x(\hy)$ is given by $N\times \gcd(N,L_y)$. \par
The similar consideration shows that there are $\gcd(N,L_x)\times \gcd(N,L_x,L_y)$ distinct loops of $W_{dp:x}(\hy)$. 
Overall, there are $N\times\gcd(N,L_x)\times\gcd(N,L_x)\times \gcd(N,L_x,L_y)$ distinct non-contractible loops of the gauge field $\hat{A}_{(xx)}$ in the $x$-direction. One can analogously count the number of the distinct non-contractible loops of $\hat{A}_{(yy)}$ in the $y$-direction, arriving at the same number. 
Therefore, the GSD, which is equivalent to the distinct number of loops of the gauge fields $\hat{A}_{(xx)}$ and $\hat{A}_{(yy)}$, is given by
\begin{equation}
    \boxed{GSD=[N\times\gcd(N,L_x)\times\gcd(N,L_x)\times \gcd(N,L_x,L_y)]^2.}\label{gsdqp}
\end{equation}
Analogously to the previous case, the foliated BF theory show the GSD dependence on the UV lattice, involving the greatest common divisor between $N$ and the system size, which is in sharp contrast with the sub-extensive GSD dependence found in the preexisting foliated BF theories.

\subsection{UV stabilizer model}
Similarly to the previous section, we can map from the BF theory to the UV lattice stabilizer model.
To start, we introduce a 2d square lattice where we 
place two types of $N$-qubit state ($\mathbb{Z}_N$ clock states) on each vertex.
We represent basis of the two types of the clock states as $\ket{a}_{(\hat{x},\hat{y})}\ket{b}_{(\hat{x},\hat{y})}$ with $a,b \in \mathbb{Z}_N$,
and the $\mathbb{Z}_N$ Pauli operators acting on the state as 
 \begin{figure}[t]
    \begin{center}
         \includegraphics[width=0.35\textwidth]{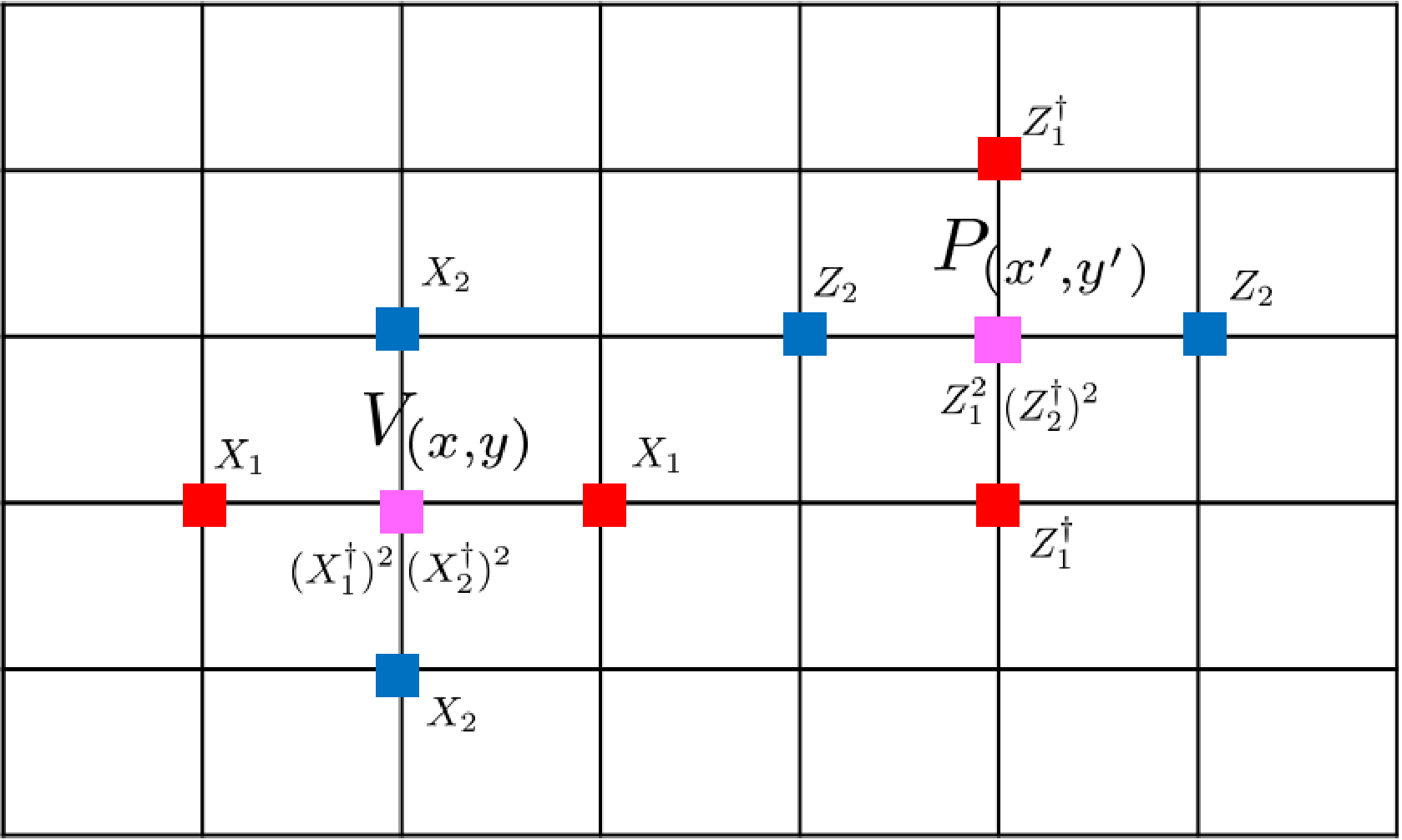}
       \end{center}
       \caption{Two types of the terms defined in~\eqref{VP}. The red and blue squares distinguish between the Pauli operators that act on the state $\ket{a}$ and those on $\ket{b}$ respectively, whereas the pink square represents the composite of the Pauli operators that act on both of $\ket{a}$ and $\ket{b}$.
 }
        \label{plane}
   \end{figure}
\begin{eqnarray}
Z_{1,(\hat{x},\hat{y})}\ket{a}_{(\hat{x},\hat{y})}\ket{b}_{(\hat{x},\hat{y})}=\omega^a\ket{a}_{(\hat{x},\hat{y})}\ket{b}_{(\hat{x},\hat{y})},\quad
Z_{2,(\hat{x},\hat{y})}\ket{a}_{(\hat{x},\hat{y})}\ket{b}_{(\hat{x},\hat{y})}=\omega^b\ket{a}_{(\hat{x},\hat{y})}\ket{b}_{(\hat{x},\hat{y})} , \nonumber\\
X_{1,(\hat{x},\hat{y})}\ket{a}_{(\hat{x},\hat{y})}\ket{b}_{(\hat{x},\hat{y})}=\ket{a+1}_{(\hat{x},\hat{y})}\ket{b}_{(\hat{x},\hat{y})},\quad
X_{2,(\hat{x},\hat{y})}\ket{a}_{(\hat{x},\hat{y})}\ket{b}_{(\hat{x},\hat{y})}=\ket{a}_{(\hat{x},\hat{y})}\ket{b+1}_{(\hat{x},\hat{y})}.
\end{eqnarray}
With this preparation, we introduce the following terms by (see Fig.~\ref{plane})
\begin{eqnarray}
V_{(\hat{x},\hat{y})}
&\vcentcolon=& X_{1,(\hat{x}+1,\hat{y})}X_{1,(\hat{x}-1,\hat{y})}(X_{1,(\hat{x},\hat{y})}^{\dagger})^2X_{2,(\hat{x},\hat{y}+1)}X_{2,(\hat{x},\hat{y}-1)}(X_{2,(\hat{x},\hat{y})}^{\dagger})^2 , \nonumber\\
P_{(\hat{x},\hat{y})}
&\vcentcolon=& Z_{1,(\hat{x},\hat{y}+1)}^{\dagger}Z_{1,(\hat{x},\hat{y}-1)}^{\dagger}Z^2_{1,(\hat{x},\hat{y})}Z_{2,(\hat{x}+1,\hat{y})}Z_{2,(\hat{x}-1,\hat{y})}(Z_{2,(\hat{x},\hat{y})}^{\dagger})^2 ,
\label{VP}
\end{eqnarray}
Note that $V_{(\hat{x},\hat{y})}$ and $P_{(\hat{x},\hat{y})}$ is what corresponds to the Gauss law and flux operator in the 
BF theory~\eqref{foliation qupole}, where it reads as $G=\Delta_x^2E_{xA}+\Delta_y^2E_{yA}$, $F_A= \Delta_x^2\hat{A}_{(yy)}-\Delta_y^2\hat{A}_{(yy)}$, respectively. 
It is straightforward to check that these operators mutually commute. 
Hamiltonian is defined by
\begin{equation}
    H_{qp}=-\sum_{\hat{x},\hat{y}} \left[ V_{(\hat{x},\hat{y})}+P_{(\hat{x},\hat{y})} \right]  +({\rm h.c. }) . 
    \label{zn}
\end{equation}

This model is exactly solvable as individual term in the Hamiltonian commutes with one another. Further, the model exhibits unusual behaviour of the ground state degeneracy on torus geometry, depending on the system size~\cite{PhysRevB.107.125154}. Differing the details to the Appendix~\ref{ap:lattice}, we derive the GSD of the stabilizer model, arriving at the same value as~\eqref{gsdqp}.

\subsection{Multipole one-form symmetries}
\label{4.4}
 \begin{figure}[t]
    \begin{center}
    \begin{subfigure}[h]{0.23\textwidth}
  \includegraphics[width=\textwidth]{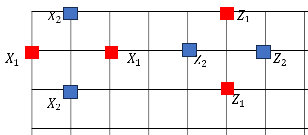}
         \caption{}\label{zw}
             \end{subfigure}
             \hspace{5mm}
       \begin{subfigure}[h]{0.23\textwidth}
  \includegraphics[width=\textwidth]{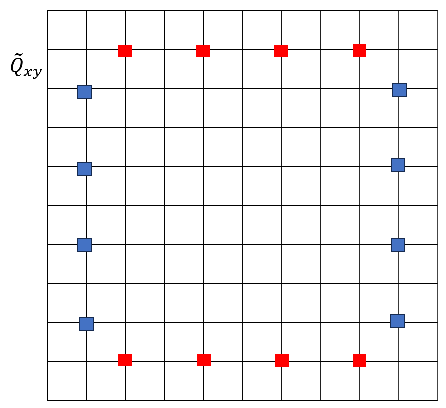}
         \caption{}\label{even}
             \end{subfigure}
             \hspace{5mm}
               \begin{subfigure}[h]{0.23\textwidth}
    \includegraphics[width=\textwidth]{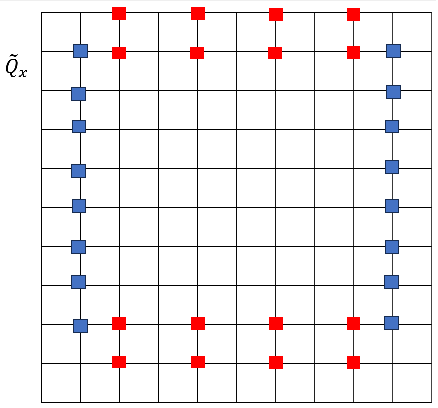}
         \caption{}\label{odd}
             \end{subfigure} 
                    \begin{subfigure}[h]{0.23\textwidth}
  \includegraphics[width=\textwidth]{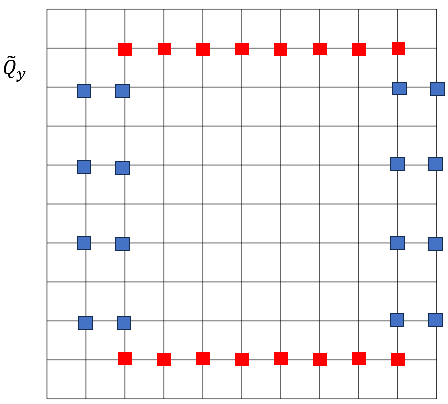}
         \caption{}\label{even1}
             \end{subfigure}
             \hspace{5mm}
               \begin{subfigure}[h]{0.23\textwidth}
    \includegraphics[width=\textwidth]{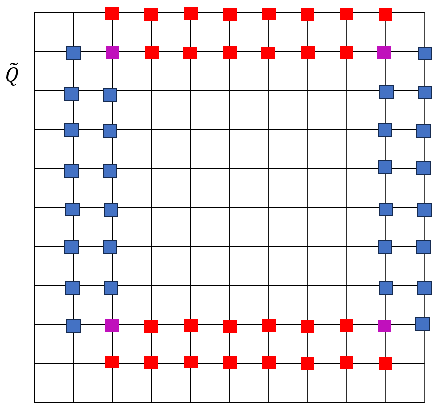}
         \caption{}\label{odd1}
             \end{subfigure} 
 \end{center}
 \caption{(a) Two type of the terms $V_{(\hx,\hy )}$ and $P_{(\hx,\hy)}$ constituting Hamiltonian~\eqref{zn} with $N=2$. (b)-(d) Four closed loops of the operators $Z_1$'s and $Z_1$'s that commute with the Hamiltonian. The red and blue squares distinguish between the Pauli operators that act on the state $\ket{a}$ and those on $\ket{b}$.}
 \end{figure}
It is known that in the $2+1d$ toric code, there are $1$-form symmetries~\cite{kapustin2014coupling}, 
which are prototype examples of the generalized global symmetries~\cite{Gaiotto:2014kfa}. In condensed matter physics language, the $1$-form symmetries in the toric code
are nothing but the closed loops of the fractionalized charges on the lattice.
Since our UV lattice model~\eqref{zn} has the simple form, one can explicitly see that 
there are such $1$-form symmetries, corresponding to closed loops. 
In addition, such symmetries follow the similar relation of the multipole symmetries that we started with~\eqref{eq:algebra}, which is not seen in the regular toric code.\par
To this end, we focus on the case with $N=2$ while other $N$ cases can be argued in a similar way. 
In this case, two type of the terms $V_{\hx,\hy}$ and $P_{(\hx,\hy)}$ constituting the Hamiltonian~\eqref{zn} becomes~(Fig.~\ref{zw})
\begin{eqnarray}
V_{(\hx,\hy)} &=& X_{1,(\hx+1,\hy)}X_{1,(\hx-1,\hy)}X_{2,(\hx,\hy+1)}X_{2,(\hx,\hy-1)} , \nonumber\\
P_{(\hx,\hy)} &=& Z_{1,(\hx,\hy+1)}Z_{1,(\hx,\hy-1)}Z_{2,(\hx+1,\hy)}Z_{2,(\hx-1,\hy)}.\nonumber\\
&&
\label{VP2}
\end{eqnarray}
There are four types of closed loops of the operators $Z_1$'s and $Z_2$'s that commute with the Hamiltonian, as portrayed in Fig.~\ref{even}-\ref{odd1}.
The first loop~(Fig.~\ref{even}), denoted by $\tilde{Q}_{xy}$, consists of two horizontal lines of $Z_1$'s and two vertical lines of $Z_2$'s, where each $Z_1$ ($Z_2$) is separated by a single site in the horizontal (vertical) direction. 
The second loop~(Fig.~\ref{odd}), labeled by $\tilde{Q}_{x}$, is formed in such a way that there are four horizontal lines of~$Z_1$'s, each of which is separated by a single site, and there are two vertical lines of~$Z_2$'s. 
The role of $Z_1$'s and $Z_2$'s is switched in the third loop, $\tilde{Q}_y$ (Fig.~\ref{even1}). 
The fourth loop, $\tilde{Q}$ consists of four horizontal lines of $Z_1$'s and four vertical lines of $Z_2$'s. We consider translating these operators by one lattice site in either $x$- or $y$-direction, and see how these loops change. 
Denoting a transnational operator in the $I$-direction by one lattice site by $T_I$, 
we investigate the change of the translation by taking the ratio between a loop after the translation and the one before the translation. We have
\begin{eqnarray}
    \frac{T^{-1}_x\tilde{Q}_{xy}T_x}{\tilde{Q}_{xy}}=\tilde{Q}_{y}, &\quad&  \frac{T^{-1}_y\tilde{Q}_{xy}T_y}{\tilde{Q}_{xy}}=\tilde{Q}_{x}, \nonumber\\ \frac{T^{-1}_x\tilde{Q}_{x}T_x}{\tilde{Q}_{x}}=\tilde{Q},&\quad&
    \frac{T^{-1}_y\tilde{Q}_{y}T_y}{\tilde{Q}_{y}}=\tilde{Q}.
    \label{1form}
\end{eqnarray}
This relation is inherited from our original consideration of a theory with multipole symmetries~\eqref{eq:algebra} in the sense that if we translate a loop, it gives rise to another loop, similarly to the fact that translating a quadruple gives a dipole (or translating a dipole gives a charge) in~\eqref{eq:algebra}. 
It would be interesting to study these $1$-form symmetries from the perspective higher-form gauging and anomaly. Also, since these $1$-form symmetries are qualitatively different from those found in the regular toric code due to~\eqref{1form}, it could be intriguing to address whether the lattice model~\eqref{zn} has different features in view of quantum error corrections~\cite{dennis2002topological}. 
We will leave these issues for future works.

\section{Conclusion and discussions}
\label{sec5}
Symmetry has been one of the fundamental laws of physics, and is a guiding principle, allowing us to 
analyze various physical properties. 
In this work, we have demonstrated that symmetry plays a pivotal role to construct new field theories in the context of the fracton topological phases. 
We have demonstrated a way to construct the $2+1d$ foliated BF theories based on the argument of the multipole symmetries. 
By introducing gauge fields associated with these symmetries, one can systematically construct new foliated BF theories. 
These theories exhibit unusual GSD dependence on the system size, involving the greatest common divisor between $N$ and the length of the lattice. 
We have also shown that these foliated BF theories are effective field theory description of unconventional topological phases, referred to as the higher rank topological phases. 
Our consideration provides a unified insight on various unconventional topological phases such as fracton and higher rank topological phases in terms of the foliated topological field theories.
\par
There are several future research directions regarding the present work that we would like to pursue for future studies. In this paper, we focus on the construction of the $2+1d$ foliated BF theories, which can be generalized in a several ways. For instance, one could study the construction of the $3+1$d or even higher dimensional foliated BF theories, with various multipole symmetries higher than dipole or quadrupole, such as ocutopoles. Also, one could think of higher-form symmetries analogue of the multipole symmetries in the same way as we did in this paper and see what the resulting foliated BF theories are.
Studying a boundary theory and topological defects of our foliated BF theories to see how multipole degrees of freedom are incorporated could be an interesting direction. 
\par
It would be also interesting to study whether our model can be useful in the context of quantum computing. 
The UV stabilizer lattice models that we have discussed here are distinct from the conventional toric code due to the sensitivity to the local geometry. 
Also, as seen from Sec.~\ref{4.4}, the model admits $1$-form symmetries which have the similar commutation relation with the transnational operators as the ones of the multipole symmetries that we started with.
In the case of the toric code, it is known that the capability of the quantum error correction is characterized by the classical Ising universal class~\cite{dennis2002topological}. 
Since the UV lattice model that we discuss respects the multipole symmetries, one naively would expect that robustness of the model against errors is qualitatively different (See~\cite{PhysRevLett.123.230503} for relevant discussion.). 
It would be intriguing to investigate whether the UV lattice models discussed in the present paper show
different universality classes that characterize the capability of the error correction.

\section*{Acknowledgement}
We thank Biswarup Ash, Bo Han, Rohit Kalloor, Ken Shiozaki, Masataka Watanabe for discussions. 
H.~E. is supported by KAKENHI-PROJECT-23H01097.
This research was conducted while M.~H. visited the Okinawa Institute of Science and Technology (OIST) through the Theoretical Sciences Visiting Program (TSVP).
M.~H. is supported by MEXT Q-LEAP, JSPS Grant-in-Aid 
for Transformative Research Areas (A) ``Extreme Universe" JP21H05190 [D01] and JSPS KAKENHI Grant Number 22H01222.
M.~H. and T.~N. are supported by JST PRESTO Grant Number JPMJPR2117.
T.~N is supported by JST, the establishment of university fellowships towards the creation of science technology innovation, Grant Number
JPMJFS2123.
\bibliographystyle{apsrev}
\bibliography{fracton-SSB}

\begin{thebibliography}{81}
\bibitem[{foo({\natexlab{a}})}]{foot3}
\bibinfo{note}{Our foliated BF theory is the effective field theory description
  of the~$\mathbb{Z}_N$ Higgs phase of the symmetric tensor gauge theory.}
\bibitem[{Tec()}]{TechnicalCaveat}
\bibinfo{note}{One can evaluate the non-contractible Wilson loops of the gauge
  fields, $\hat{A}_{(ij)}$ and its canonical conjugate,
  $(\hat{c}^y_x,\hat{c}^x_y,\tilde{\hat{c}})$ to count the number of the
  Heisenberg algebras to identify the GSD. Since this task is complicated as
  the Wilson loops of the both of the gauge fields, $\hat{A}_{(ij)}$ and
  $(\hat{c}^y_x,\hat{c}^x_y,\tilde{\hat{c}})$ highly depend on the spatial
  coordinate. Here, we employ an alternative simpler approach to obtain the GSD
  by focusing on the Wilson loops of the gauge field, $\hat{A}_{(ij)}$ and
  count the number of distinct configurations of the loops.}
\bibitem[{foo({\natexlab{b}})}]{foot2}
\bibinfo{note}{Again, we put the theory on the discrete 2d lattice with
  coordinate ($\hx,\hy$), and due to the periodic boundary condition,
  $(\hx+L_x,\hy)\sim(\hx,\hy+L_y)\sim(\hx,\hy)$.}
\end{thebibliography}

\appendix
\begin{widetext}
\section{Mapping between real BF and integer BF theory}
\label{app1}
Generally the BF theory with multipole symmetry studied in the present paper contains the higher order spatial derivatives, making analysis much more challenging compared with the conventional BF theory of the topologically ordered phases. 
To circumvent this problem, we make use of a mapping from the BF theory to  what is called integer BF theory where gauge fields take integer values defined on a discrete lattice, proposed in~\cite{Gorantla:2021svj}.
The latter BF theory is especially useful to study the properties of the model with multipole symmetries, such as the Wilson loops.
For clearer illustration purposes, we first review how such a mapping works in the case of the toric code and 
then we apply this mapping to our BF theory.\par
Let us introduce the following BF theory defined on a discrete infinite Euclidean lattice:
\begin{equation}
    L=\frac{2\pi}{N}\biggl[\hat{a}_{\tau}(\Delta_x \hat{b}_x-\Delta_y \hb_x)+\hat{a}_{x}(\Delta_y \hat{b}_{\tau}-\Delta_{\tau} \hb_y)+\hat{a}_{y}(\Delta_{\tau} \hat{b}_x-\Delta_x \hb_{\tau})\biggr].\label{integer}
\end{equation}
Here, gauge fields $\ha_{\mu}$ $\hb_{\mu}$ ($\mu=\tau,x,y$) take integer values defined on a $\mu$-link of the infinite Euclidean lattice. Following the terminology in~\cite{Gorantla:2021svj,hotan2022,PhysRevB.106.045112}, we term such a theory integer BF theory. 
This integer BF theory corresponds to the BF theory of the toric code~\eqref{toric code}. 
To see why, we first think of the equivalent description of~\eqref{integer} which is given by
\begin{eqnarray}
    L=\frac{N}{2\pi}\biggl[a_{\tau}(\Delta_x b_x-\Delta_y b_x-2\pi m_{xy})+a_{x}(\Delta_y b_{\tau}-\Delta_{\tau} b_y-2\pi m_{y\tau})+a_{y}(\Delta_{\tau} b_x-\Delta_x b_{\tau}-2\pi m_{\tau x})\biggr]\nonumber\\
    -Nn_{xy}b_{\tau}- Nn_{\tau y}b_{x} -Nn_{\tau x}b_{y}
    +n_{xy}\Delta_{\tau}\tilde{\phi}+ n_{ y\tau}\Delta_x\tilde\phi+n_{\tau x}\Delta_y\tilde{\phi} +m_{xy}\Delta_{\tau}\phi+ m_{ y\tau}\Delta_x\phi+m_{\tau x}\Delta_y\phi.\label{as}
\end{eqnarray}
Here, $a_{\mu}$ $b_{\mu}$ ($\mu=\tau,x,y$) represent gauge fields with real values. Note the distinction between the fields with or without hat. We put hat on top of the fields to emphasize that they are the fields with integer values.
Further, we have intruded the Stueckelberg fields, $\phi$, $\tilde\phi$ to ensure that $a_{\mu}$ $b_{\mu}$ respects the gauge symmetry and 
$m_{\mu\nu}$, $n_{\mu\nu}$ as the integer fields defined on a plaquette in the $\mu\nu$-plane, which take the role of the Lagrangian multiplier. 
Indeed, summing over the integer fields $m_{\mu\nu}$, $n_{\mu\nu}$ gives the following constraints
\begin{equation}
    a_\mu=\frac{2\pi}{N}\ha_u+\frac{1}{N}\Delta_{\mu}\phi,\;\; b_\mu=\frac{2\pi}{N}\hb_u+\frac{1}{N}\Delta_{\mu}\tilde\phi,\label{hey}
\end{equation}
where $\ha_{\mu}$ $\hb_{\mu}$ ($\mu=\tau,x,y$) are integer fields. Substituting~\eqref{hey} to~\eqref{as}, we come back to \eqref{integer}. 
The theory~\eqref{as} admits the following gauge symmetry:
\begin{eqnarray}
&&   a_\mu\to a_\mu+\Delta_{\mu}\alpha+2\pi \hat{k}_{\mu},\quad  b_\mu\to b_\mu+\Delta_{\mu}\beta+2\pi \hat{q}_{\mu},\quad
\phi\to\phi+N\alpha+2\pi \hat{k}_{\phi},\quad
\tilde\phi\to\tilde\phi+N\beta+2\pi \hat{q}_{\tilde\phi},
     \nonumber\\
&&     m_{xy}\to m_{xy}+\Delta_{x}\hat{k}_y-\Delta_y\hat{k}_x,\quad 
m_{y\tau }\to m_{y\tau}+\Delta_{y}\hat{k}_{\tau}-\Delta_\tau\hat{k}_{y},\quad 
m_{\tau x}\to m_{\tau x}+\Delta_{\tau}\hat{k}_x-\Delta_x\hat{k}_{\tau}\nonumber\\
&&      n_{xy}\to n_{xy}+\Delta_{x}\hat{q}_y-\Delta_y\hat{q}_x,\quad 
n_{y\tau }\to n_{y\tau}+\Delta_{y}\hat{q}_{\tau}-\Delta_\tau\hat{q}_{y},\quad 
n_{\tau x}\to n_{\tau x}+\Delta_{\tau}\hat{q}_x-\Delta_x\hat{q}_{\tau}.\label{gauge sym}
\end{eqnarray}
Here, $\alpha,\beta$ represent the real gauge parameters, whereas $\hat{k}_{\mu},\hat{q},\hat{k}_{\phi},\hat{q}_{\tilde\phi}$ denote the integer gauge parameters.
To see the relation to the original BF theory of the toric code~\eqref{toric code}, we sum over the Stueckelberg fields, $\phi$ $\tilde{\phi}$, in~\eqref{as} which gives the following flux-less condition:
\begin{equation}
    \Delta_{\tau}m_{xy}+\Delta_x m_{y\tau}+\Delta_ym_{\tau x}=0\label{fluxl}
\end{equation}
and similarly for $n_{\mu\nu}$. Since we think of the infinite Euclidean lattice, using the gauge symmetry~\eqref{gauge sym}, jointly with the flux-less condition~\eqref{fluxl} allows us to set the fields 
$m_{\mu\nu}$ $n_{\mu\nu}$ to be zero. 
(In the case of the lattice with periodic boundary condition, $m_{\mu\nu}$ $n_{\mu\nu}$ can be set to be zero except a few cells, which capture the holonomy. 
Accordingly, one needs to take into account the transition function of the gauge fields, $a_\mu$, $b_{\nu}$, see~\cite{Gorantla:2021svj} for more discussion on this point.)
What remains in~\eqref{as} is 
\begin{equation}
    L=\frac{N}{2\pi}\biggl[a_{\tau}(\Delta_x b_x-\Delta_y b_x)+a_{x}(\Delta_y b_{\tau}-\Delta_{\tau} b_y)+a_{y}(\Delta_{\tau} b_x-\Delta_x b_{\tau})\biggr].\label{dd}
\end{equation}
After taking the continuum limit appropriately, \eqref{dd} becomes the original BF theory of the toric code~\eqref{toric code}.\par
After having reviewed the mapping between real BF theory and the integer BF theory, now we apply this technique to our theories. In Sec.~\ref{sec3}, we considered the BF theory with dipole symmetry given in~\eqref{Bfdi}. 
Since the way we carry out the a mapping between theories closely parallels the one in the case of the toric code, we succinctly describe how the mapping works.
The corresponding integer BF theory defined on the infinite discrete Euclidean lattice is given by
\begin{eqnarray}
    \mathcal{L}_{dip}
   &=&\frac{2\pi}{N}\biggl[ 
   -\hat{c}_0^x \left( \Delta_x\hat{A}_{(yx)}-\Delta_y\hat{A}_{(xx)} \right)
   -\hat{c}_0^y \left( \Delta_x\hat{A}_{(yy)}-\Delta_y\hat{A}_{(xy)} \right)
 +\hat{\tilde{c}}\left( \Delta_{\tau}\hat{A}_{(xy)}-\Delta_x\Delta_y\hat{A}_0 \right) \nonumber\\
  &&
  -\hat{c}_y^x \left( \Delta_{\tau}\hat{A}_{(xx)}-\Delta_x^2\hat{A}_0 \right)+\hat{c}_x^y \left( \Delta_{\tau}\hat{A}_{(yy)}-\Delta_y^2\hat{A}_0 \right )\biggr],\label{ppq}
\end{eqnarray}
where the gauge fields with hat~($\hat{\cdot}$) take integer values.
To see such a mapping works, let us first rewrite~\eqref{ppq} as
\begin{eqnarray}
    \mathcal{L}_{dip}
   &=&\frac{N}{2\pi}\biggl[ 
   -c_0^x \left( \Delta_xA_{(yx)}-\Delta_yA_{(xx)}-2\pi m^x_{xy} \right)
   -{c}_0^y \left( \Delta_xA_{(yy)}-\Delta_yA_{(xy)}-2\pi m^y_{xy} \right)\nonumber\\
   &+&\tilde{c}\left( \Delta_{\tau}A_{(xy)}-\Delta_x\Delta_yA_0 -2\pi \tilde{m}\right)
   -c_y^x \left( \Delta_{\tau}A_{(xx)}-\Delta_x^2A_0-2\pi m^x_{\tau x} \right)+c_x^y \left( \Delta_{\tau}A_{(yy)}-\Delta_y^2A_0 -2\pi m^y_{y\tau}\right )\biggr]\nonumber\\
   &-&NA_0n_{xy}-NA_{(xy)}\tilde n-NA_{(xx)}n_{y\tau}-NA_{(yy)}n_{\tau x}-m^x_{xy}\Delta_{\tau}\phi^x-m^y_{xy}\Delta_\tau\phi^y\nonumber\\
   &+&\tilde{m}(\Delta_x\phi^x-\Delta_y\phi^y)+m_{\tau x}\Delta_y\phi^x+n_{xy}\Delta_\tau\tilde\phi+\tilde\Delta_x\Delta_y\tilde\phi+n_{y\tau}\Delta_x^2\tilde{\phi}+n_{\tau x}\Delta_y^2\tilde{\phi}.\label{a8}
\end{eqnarray}
Here, $(c_0^x,c_0^y,c^x_y,c^y_x,\tilde{c})$, $(A_0,A_{(xx)},A_{yy},A_{(xy)})$ denote the gauge fields which take \textit{real} values with the Stueckelberg fields $\phi^x,\phi^y,\tilde\phi$ to ensure the gauge symmetry. Also, integer fields $m^i_{\mu\nu},\tilde{m},n^i_{\mu\nu},\tilde n$ ($\mu,\nu=x,y,\tau$, $i=x,y$) are Lagrangian multipliers; summing over these fields yields
\begin{eqnarray}
&&    c^i_0=\frac{2\pi}{N}\hat{c}_0^i+\frac{1}{N}\Delta_i\phi^i\ \ (i=x,y),\quad
c_x^y=\frac{2\pi}{N}\hat{c}_x^y+\frac{1}{N}\Delta_x\phi^y, \nonumber\\
&& c_y^x=\frac{2\pi}{N}\hat{c}_y^x+\frac{1}{N}\Delta_y\phi^x, \quad
\tilde c=\frac{2\pi}{N}\hat{\tilde{c}}+\frac{1}{N}\left( \Delta_x\phi^x-\Delta_y\phi^y \right) \nonumber\\
&&    A_0=\frac{2\pi}{N}\hat{A}_0+\frac{1}{N}\Delta_\tau\tilde\phi, \quad
A_{(ij)}=\frac{2\pi}{N}\hat{A}_{(ij)}+\frac{1}{N}\Delta_i\Delta_j\tilde\phi,
\end{eqnarray}
where the fields with hat represent the integer fields.
One can easily check that substituting this into~\eqref{a8} gives~\eqref{ppq}. 
Similarly to the previous case of the toric code, when
we sum over the Stueckelberg fields $\phi^x,\phi^y,\tilde\phi$ gives the flux-less condition of the fields $m^i_{\mu,\nu},\tilde{m},n^i_{\mu\nu},\tilde n$. 
Using this condition, jointly with the gauge symmetries of these fields allows us to suppress the fields $m^i_{\mu,\nu},\tilde{m},n^i_{\mu\nu},\tilde n$. What remains in~\eqref{a8} turns out to be
\begin{eqnarray}
    \mathcal{L}_{dip}
   &=&\frac{N}{2\pi}\biggl[ 
   -c_0^x \left( \Delta_xA_{(yx)}-\Delta_yA_{(xx)}\right)
   -{c}_0^y \left( \Delta_xA_{(yy)}-\Delta_yA_{(xy)} \right)\nonumber\\
   &+&\tilde{c}\left( \Delta_{\tau}A_{(xy)}-\Delta_x\Delta_yA_0\right)
   -c_y^x \left( \Delta_{\tau}A_{(xx)}-\Delta_x^2A_0\right)+c_x^y \left( \Delta_{\tau}A_{(yy)}-\Delta_y^2A_0\right )\biggr].
\end{eqnarray}
Taking the continuum limit, we obtain the original BF theory~\eqref{Bfdi}. Based on this, to study the original BF theory~\eqref{Bfdi}, we instead study the integer BF theory defined on the discrete lattice~\eqref{ppq}, which allows us to study its physical properties. \par
By the analogous lines of thoughts, we study the integer BF theory~\eqref{eq:bf0} with quadrupole instead of the BF theory introduced in~\eqref{eq:bf} in Sec.~\ref{sec4}.

\section{Derivation of~\eqref{dpgsd}}
\label{ap:1}

In this appendix, we provide an argument to derive \eqref{dpgsd}.
The composite of loops~\eqref{comosite} is subject to the conditions
\begin{equation}
    (k_x,k_y)\sim (k_x+N,k_y)\sim (k_x+\alpha_y L_y,k_y)\label{a1}
\end{equation}
and 
\begin{equation}
      (k_x,k_y)\sim (k_x,k_y+N)\sim (k_x,k_y+\alpha_x L_y)
\end{equation}
with $\alpha_{x/y}=\frac{N}{\gcd(N,L_{x/y})}$.
Also, $(k_x,k_y)$ has to satisfy
\begin{eqnarray}
  (k_x,k_y)\sim  (k_x+L_x,k_y-L_y).\label{a3}
\end{eqnarray}
From~\eqref{a1} and \eqref{a3}, there are $ \gcd(N,\alpha_yL_x,L_x)$
distinct number of $k_x$ and for these distinct values of $k_x$, there are $\gcd(N,\alpha_xL_y)$ distinct number of $k_y$. In total, there are $\gcd(N,\alpha_yL_x,L_x)\times  \gcd(N,\alpha_xL_y)$ distinct number of the composite loops~\eqref{comosite}. 
Since $\gcd(N,\alpha_y L_x,L_x)=\gcd(N,L_x)$,
it follows that
\begin{eqnarray}
      &&\gcd(N,\alpha_yL_x,L_x)\times  \gcd(N,\alpha_xL_y)\nonumber\\
      &&=\gcd(N,L_x)\times \gcd\biggl(N,\frac{N}{\gcd(N,L_x)}L_y\biggr)\nonumber\\
      &&= N\gcd(\gcd(L_x,N),L_y)=N\gcd(N,L_x,L_y).\nonumber\\
      &&
\end{eqnarray}
Hence, there are $N\gcd(N,L_x,L_y)$ distinct loops of the gauge field~$\hat{A}_{(xy)}$. Taking into the other non-contractible loops into account, we arrive at that the GSD, which is the number of the distinct non-contractible loops of the gauge fields $\hat{A}_{(ij)}$ is given by~\eqref{dpgsd}.

\section{Investigation of the lattice model by the Laplacian}
\label{ap:lattice}
In this appendix, we analyze the stabilizer model~\eqref{zn} obtained from the foliated BF theory, especially investigate the Wilson loops that contributes to the non-trivial GSD on a torus geometry from a different perspective.  
As we discussed in the main text, the model contains the second order spatial derivatives, involving the nearest neighboring states. Due to this property, one can make use of the \textit{Laplacian}, which is the graph theoretical analogue of the second order derivatives~\cite{chung1997spectral}, allowing us to systematically investigate the lattice model, such as the Wilson loops. To see how, we first recall the Hamiltonian has the following form, 
\begin{equation}
     H_{qp}=-\sum_{\hat{x},\hat{y}} \left[ V_{(\hat{x},\hat{y})}+P_{(\hat{x},\hat{y})} \right] +({\rm h.c.}),
     \label{c1}
\end{equation}
and that the ground state~$\ket{\Omega}$ satisfies $V_{(\hx,\hy)}\ket{\Omega}=P_{(\hx,\hy)}\ket{\Omega}=\ket{\Omega}$, i.e., the ground state does not admit an electric and magnetic excitation. 
To discuss the Wilson loops, we also investigate the excitations.
When acting an operator $Z_{1,(\hx,\hy)}$ on the ground state, it violates the condition $V_{(\hx,\hy)}=1$, that is,
\begin{eqnarray}
    &&V_{(\hx,\hy)}(Z_{1,(\hx,\hy)}\ket{\Omega})=\omega^{-2}(Z_{1,(\hx,\hy)}\ket{\Omega}),\nonumber\\
    &&V_{(\hx\pm1,\hy)}(Z_{1,(\hx,\hy)}\ket{\Omega})=\omega (Z_{1,(\hx,\hy)}\ket{\Omega})\label{fusion}
\end{eqnarray}
yielding electric charges. This implies that by acting an operator $Z_{1,(\hx,\hy)}$, an electric charge is induced at the coordinate $(\hx\pm1,\hy)$ and two conjugate of the electric charges are obtained at $(\hx,\hy)$.
Denoting the $\mathbb{Z}_{N}$ eclectic charge as $e_{(\hx,\hy)}$ (and its conjugate as $\overline{e}_{(\hx,\hy)}$), 
\eqref{fusion} can be described by the fusion rule, reading
\begin{equation}
    I\ \to\ e_{(\hx-1,\hy)}\overline{e}^2_{(\hx,\hy)}e_{(\hx+1,\hy)}.\label{fusion2}
\end{equation}
The fusion rule of the magnetic charges can be similarly discussed.
\par
Using this property, one can study the GSD of the model on a torus geometry by counting the distinct number of Wilson loops. To this end, the Laplacian comes into play.
We consider the model~\eqref{c1} on a 2d lattice with periodic boundary condition and the system size being $L_x\times L_y$. In this lattice, we think of a closed loop of the electric charge in the horizontal direction at $\hy$, described by $\prod_{\hx=1}^{L_x}Z_{1,(\hx,\hy)}^{a_{\hx}}$ with $a_{\hx}\in\mathbb{Z}_N$.
From~\eqref{fusion2}, the electric charges induced by acting this operator on the ground state are described by the fusion rule
\begin{equation}
    I\to \prod_{\hx=1}^{L_x}\otimes e^{r_{\hx}}_{(\hx,\hy)}\quad
    (r_{\hx}\in\mathbb{Z}_N)
    \label{b4}
\end{equation}
with 
\begin{equation}
    \mathbf{r}=-L_{{L_x\times L_x}}\mathbf{a}.\label{b5}
\end{equation}
Here, $\mathbf{a}\vcentcolon=(a_1,\cdots,a_{L_x})^T$, $\mathbf{r}\vcentcolon=(r_1,\cdots,r_{L_x})^T$, and $L_x\times L_x$ matrix, $L_{{L_x\times L_x}}$, is the Laplacian, the graph theoretical analogue of the second order spatial derivative, which has the following form
\begin{equation}
         L_{L_x\times L_x}=\begin{pmatrix}
2 & -1&&& -1\\
-1 & 2&-1 &&\\
&-1&2&\ddots&\\
&&\ddots&\ddots&-1\\
-1&&&-1&2\label{lp}
\end{pmatrix}.
\end{equation}
Since the closed non-contractible loop~$\prod_{\hx=1}^{L_x}Z_{1,(\hx,\hy)}^{a_{\hx}}$ has to commute with the Hamiltonian, the fusion rules~\eqref{b4} and \eqref{b5} have to be trivial, i.e., $\mathbf{r}=\mathbf{0}\mod N$. 
Thus, the closed loops of the electric charges are characterized by the \textit{kernel} of the Laplacian. \par
By evaluating the Laplacian~\eqref{lp}, the solution of $L_{L_x\times L_x}\mathbf{a}=\mathbf{0}$ is given by~\cite{PhysRevB.107.125154}
\begin{equation}
    \mathbf{a}=\alpha_0(1,1,\cdots,1)^T+\alpha_1\times\frac{N}{\gcd(N,L_x)}(1,2,3,\cdots, L_x )^T.
\end{equation}
\begin{figure}[t]
    \begin{center}
       \begin{subfigure}[h]{0.45\textwidth}
  \includegraphics[width=\textwidth]{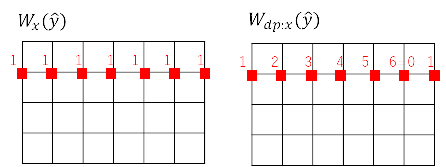}
         \caption{}\label{dip1}
             \end{subfigure}
             \hspace{5mm}
               \begin{subfigure}[h]{0.23\textwidth}
    \includegraphics[width=\textwidth]{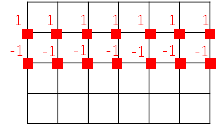}
         \caption{}\label{dip2}
             \end{subfigure} 
 \end{center}
 \caption{(a)~Examples of the two types of the loops given in~\eqref{sol} in the case of $N=L_x=6$. The red dots denote the operators $Z_{1,(\hx,\hy})$. The integer numbers correspond to the power entering in the form of the loops~~\eqref{sol}. The periodic boundary condition is imposed so that the left and right edges are identified. (b) Configuration corresponding to the second term of~\eqref{c14}, which is the composite of the loops with opposite charge located adjacent in the~$y$-direction. }
 \end{figure}
Here, $\alpha_0\in\mathbb{Z}_N$ and $\alpha_1\in \mathbb{Z}_{\gcd(N,L_x)}$. Hence, there are two kinds of the closed loops in the horizontal direction ($\alpha_x=\frac{N}{\gcd(N,L_x)}$):
\begin{equation}
    W_x(\hy)=\prod_{\hx=1}^{L_x}Z_{1,(\hx,\hy)},\quad   
    W_{dp:x}(\hy)=\prod_{\hx=1}^{L_x}( Z^{\hx}_{1,(\hx,\hy)})^{\alpha_x}.
    \label{sol}
\end{equation}
The first loop is interpreted as the conventional Wilson loop (Fig.~\ref{dip1} left), which is formed by the trajectory of the $\mathbb{Z}_N$ electric charge, traveling around the torus in the $x$-direction. This corresponds to the first term of~\eqref{418} in the BF theory. The second loop can be interpreted as the ``dipole of the Wilson loop'' (Fig.~\ref{dip1} right) in the sense that the loops is formed by the $\mathbb{Z}_N$ electric charge going around the torus with its intensity increasing linearly. This corresponds to the second term in~\eqref{418}.
The first loop, $W_x(\hy)$ is labeled by $\mathbb{Z}_N$ whereas the second one  $W_{dp:x}(\hy)$ is by $\mathbb{Z}_{\gcd(N,L_x)}$. By evaluating the fusion rules, one can similarly discuss the configuration of the (part of) closed loops in the lattice model~\eqref{op} obtained from the foliated BF theory with dipole symmetry~\eqref{foliation dipole}. 
For instance, the form of the loops given in~\eqref{loops} can be similarly derived. \par
Having identified loops of the electric charges in the $x$-direction~\eqref{sol}, one need to check whether these loops are $\hy$ dependent or not to count the distinct number of configuration of the loops. Let us focus on the loop~$W_x(\hy)$. By multiplying the following operator
\begin{equation}
    M(\hy)=\prod_{\hx=1}^{L_x}P_{(\hx,\hy)},
\end{equation}
where $P_{(\hx,\hy)}$ is given in~\eqref{VP},
the loop~$W_x(\hy)$ is deformed as
\begin{equation}
    M(\hy)W_x(\hy)=W_x(\hy-1)W_x^{\dagger}(\hy)W_x(\hy+1).\label{df}
\end{equation}
We need to count the distinct configuration of the loop~$W_x(\hy)$ up to the deformation~\eqref{df}.
To this end, we think of deforming the composite of the loops, $\prod_{\hy=1}^{L_y}W^{b_{\hy}}_x(\hy)$ with $b_{\hy}\in\mathbb{Z}_N$. 
Multiplying it with the following operator
\begin{equation}
    \prod_{\hy=1}^{L_y} M^{c_{\hy}}(\hy)\quad (c_{\hy}\in\mathbb{Z}_N)
    \label{mm}
\end{equation}
and referring to~\eqref{df}, it follows that the composite of the loops after multiplying~\eqref{mm}, becomes
\begin{equation}
    \prod_{\hy=1}^{L_y}W^{\tilde{b}_{\hy}}(\hy) \quad (\tilde{b}_{\hy}\in\mathbb{Z}_N)
\end{equation}
with
\begin{equation}
    \tilde{\mathbf{b}}=\mathbf{b}-L_{L_y\times L_y}\mathbf{c}.\label{deformed}
\end{equation}
Here, $\mathbf{b}=(b_1,\cdots,b_{L_y})^T$ with other vectors $  \tilde{\mathbf{b}}, \mathbf{c}$ being similarly defined and the $L_y\times L_y$ matrix, $L_{L_y\times L_y}$  denotes the Laplacian, which has the same form as~\eqref{lp}, where the matrix size being replaced with $L_y\times L_y$.
It follows that the composite of loops, $\prod_{\hy=1}^{L_y}W^{b_{\hy}}_x(\hy)$, labeled by the vector~$\mathbf{b}$ and the ones $\prod_{\hy=1}^{L_y}W^{\tilde{b}_{\hy}}_x(\hy)$ labeled by $\tilde{\mathbf{b}}$ which are related via~\eqref{deformed} with $^{\exists}\mathbf{c}$ are identified. Therefore, 
the number of distinct configurations of the loop~$W_x(\hy)$ found to be $\mathbf{s}\vcentcolon=\mathbb{Z}_N^{L_y}/\text{Im}(L_{L_y\times L_y})$, which is the \textit{cokernel} of the Laplacian.
Note that this corresponds to the discussion around Eq.~\eqref{419} in the main text, where we count the distinct number of the loops~$W(\hy)$ subject to the condition~\eqref{419}. Namely, in~\eqref{419}, we discuss the distinct configuration of the loops up to the operation of the second order derivative $\partial_y^2$, which corresponds to the present consideration where we need to find distinct configurations of the loops up to the Laplacian, $L_{L_y\times L_y}$.\par
The cokernel is given by~\cite{PhysRevB.107.125154}
\begin{eqnarray}
    \mathbf{s}=\beta_0(0,\cdots,0,1)^T+\beta_1(0\cdots,0,1,-1)^T\label{c14}
\end{eqnarray}
with $\beta_0\in\mathbb{Z}_N$, $\beta_1\in\mathbb{Z}_{\gcd(N,L_y)}$.
Eq.~\eqref{c14} indicates that the distinct configurations of the loops are characterized by a single loop and the composite of loops (Fig.~\ref{dip2}) which have the opposite charge, located adjacent
to each other in the $y$-direction. These two types of loops are labeled by $\mathbb{Z}_N$ and $\mathbb{Z}_{\gcd(N,L_y)}$. This corresponds to the discussion around~\eqref{diloops}. \par
Overall, we have counted the distinct number of loops~$W_x(\hy)$ which is given by
$N\times\gcd(N,L_y)$. Analogous line of thoughts leads to that there are $\gcd(N,L_x)\times\gcd(N,L_x,L_y)$ distinct configurations of the loop $W_{dp:x}(\hy)$. Thus, in total, there are $N\times\gcd(N,L_y)\times\gcd(N,L_x)\times\gcd(N,L_x,L_y)$ different configurations of the loops of the electric charge. So far, 
have considered closed loops of the electric charges. Regarding the closed loops of the magnetic charges, the similar argument follows as the electric charges, thus they are labeled by the same quantum numbers. Taking it into consideration, we finally arrive at that the GSD is the same value as~\eqref{gsdqp}.


\end{widetext}
\end{document}